\documentclass[twocolumn,tighten,apj,iop]{emulateapj}
\usepackage{amsmath, amsthm, amssymb}
\usepackage{natbib}

\newcommand{\pd}[2]{\frac{\partial #1}{\partial #2}}
\newcommand{\td}[2]{\frac{d #1}{d #2}}

\newcommand{\avg}[1]{\langle #1 \rangle}
\newcommand{\sci}[2]{#1\times 10^{#2}}

\shorttitle{Integrated Neutrino Driven Wind Nucleosynthesis}
\shortauthors{}
\shortauthors{Roberts et al.}

\begin{document}
\title{Integrated Nucleosynthesis in Neutrino Driven Winds}
\affil{} \author{L. F. Roberts\altaffilmark{1}, S. E. 
Woosley\altaffilmark{1}, and R. D. Hoffman\altaffilmark{2}}
\altaffiltext{1}{Department of Astronomy and Astrophysics, University
of California, Santa Cruz, CA 95064 USA}
 \altaffiltext{2}{Computational Nuclear Physics Group, L-414, Lawrence
 Livermore National Laboratory, Livermore, CA 94550 USA}
\begin{abstract}

Although they are but a small fraction of the mass ejected in
core-collapse supernovae, neutrino-driven winds (NDWs) from nascent
proto-neutron stars (PNSs) have the potential to contribute
significantly to supernova nucleosynthesis.  In previous works, the NDW
has been implicated as a possible source of r-process and light
p-process isotopes.  In this paper we present time-dependent
hydrodynamic calculations of nucleosynthesis in the NDW which include
accurate weak interaction physics coupled to a full nuclear reaction
network.  Using two published models of PNS neutrino luminosities, we
predict the contribution of the NDW to the integrated nucleosynthetic
yield of the entire supernova. For the neutrino luminosity histories
considered, no true r-process occurs in the most basic scenario.  The
wind driven from an older $1.4 M_\odot$ model for a PNS
 is moderately neutron-rich at late times however,
and produces $^{87}$Rb, $^{88}$Sr, $^{89}$Y, and $^{90}$Zr in near
solar proportions relative to oxygen.  The wind from a more recently
studied $1.27 M_\odot$ PNS is proton-rich
throughout its entire evolution and does not contribute significantly
to the abundance of any element.  It thus seems very unlikely that the
simplest model of the NDW can produce the r-process.  At most, it
contributes to the production of the N = 50 closed shell elements and
some light p-nuclei. In doing so, it may have left a distinctive
signature on the abundances in metal poor stars, but the results are
sensitive to both uncertain models for the explosion and the masses of
the neutron stars involved.

\end{abstract}


\section{Introduction}

The site where r-process nuclei above A=90 have been synthesized
remains a major unsolved problem in nucleosynthesis theory
\citep[e.g.,][]{Arnould07}.  Historically, many possibilities have
been proposed \citep[see][]{Meyer94}, but today, there are two
principal contenders - neutron star mergers
\citep{Lattimer77,Freiburghaus99} and the NDW
\citep{Woosley94,Qian96,Hoffman97,Otsuki00,Thompson01}.  Observations
of ultra-metal-poor stars suggest that many r-process isotopes were
already quite abundant at early times in the galaxy
\citep{Cowan95,Sneden96,Frebel07}, suggesting both a primary origin
for the r-process and an association with massive stars. NDWs would
have accompanied the first supernovae that made neutron stars and,
depending upon what is assumed about their birth rate and orbital
parameters, the first merging neutron stars could also have occurred
quite early.

Both the merging neutron star model and the NDW have problems
though. In the simplest version of galactic chemical evolution,
merging neutron stars might be capable of providing the necessary
integrated yield of the r-process in the sun, but they make it too
rarely in large doses and possibly too late to be consistent with
observations \citep{Argast04}.  On the other hand, making the
r-process in NDWs requires higher entropies, shorter time-scales, or
lower electron mole numbers, $Y_e$, than have been demonstrated in any
realistic, modern model for a supernova explosion \citep[though
  see][]{Burrows06}.

Previous papers and models for nucleosynthesis in the NDW have focused
on the production of nuclei heavier than iron using either greatly
simplified dynamics \citep{Beun08,Farouqi09} or nuclear
physics \citep{Qian96,Otsuki00,Arcones07,Fischer09,Huedepohl09}.
Post processing nuclear network calculations have been performed using 
thermal histories from accurate models of the dynamics 
\citep{Hoffman97,Thompson01,Wanajo06}, but the calculations sampled only a 
limited set of trajectories in the ejecta. No one has yet combined the 
complete synthesis of a realistic NDW with that of the rest of the 
supernova.

To address this situation, and to develop a framework for testing the
nucleosynthesis of future explosion models, we have calculated
nucleosynthesis using neutrino luminosity histories taken from two PNS
calculations found in the literature \citep{Woosley94,Huedepohl09}.
This was done using a modified version of the implicit one-dimensional
hydrodynamics code Kepler, which includes an adaptive nuclear network
of arbitrary size.  This network allows for the production of both
r-process nuclei during neutron-rich phases of the wind and production
of light p-elements during proton-rich phases.  Since the results of wind
nucleosynthesis depend sensitively on the neutrino luminosities and
interaction rates \citep{Qian96,Horowitz02}, we have included accurate
neutrino interaction rates that contain both general relativistic and
weak magnetism corrections.  

The synthesis of all nuclei from carbon through lead is integrated
over the history of the NDW and combined with the yield from the rest
of the supernova, and the result is compared with a solar distribution.
If a nucleus produced in the NDW is greatly overproduced relative to
the yields of abundant elements in the rest of the supernova, there is
a problem. If it is greatly underproduced, its synthesis in the NDW is
unimportant.  If it is co-produced, the NDW may be responsible for the
galactic inventory of this element.  An important outcome of this
study are the yields expected from a ``plain vanilla'' model for the
NDW. Are there any elements that are robustly produced and thus might
be used as diagnostics of the wind in an early generation of stars?

In \S\ref{wind_physics}, we discuss the general physics of neutrino
driven winds and analytically delineate the regions in neutrino
temperature space were different modes of nucleosynthesis occur.  We
then discuss our numerical model in \S\ref{computational_method}.  In
\S\ref{results}, the results of the time dependent models are
presented.  We conclude with a discussion of how the NDW might affect
galactic chemical evolution and consider if this allows the strontium
abundance in low metallicity halo stars to be used as a tracer of
supernova fallback at low metallicity.  Additionally, we investigate
if observed abundances in SN 1987A can put constraints on late time
neutrino luminosities from PNSs.  Finally, we discuss some possible
modifications of the basic model that might improve the r-process
production. These ideas will be explored more thoroughly in 
a subsequent paper.

\section{General Concepts and Relevant Physics} \label{wind_physics}

After collapse and bounce in a core collapse supernova, a condition of
near hydrostatic equilibrium exists in the vicinity of the
neutrinospheres. The temperature of the outer layers is changing on a
time scale determined by the Kelvin-Helmholtz time of the PNS,
$\tau_{KH} \approx 10 s$ \citep{Burrows86,Pons99}.  This is much
longer than the dynamical time scale of the PNS envelope, so the
hydrostatic part of the envelope is in an approximate steady state.
The neutrino heating rate, which is determined by the neutrino
luminosities from the neutrino sphere, must then balance the local
neutrino cooling rate.  Heating and cooling are dominated by the
charged current processes $(\nu_e + n) \rightleftharpoons (e^- + p)$
and $(\bar \nu_e + p) \rightleftharpoons (e^+ + n)$ \citep{Qian96}.
Equating these rates, while neglecting the neutron-proton mass
difference and weak magnetism corrections and assuming the geometry
can be approximated as close to plane-parallel gives the temperature
structure of the neutron star atmosphere as a function of radius,
$T_{atm} \approx 1.01 \, \textrm{MeV} \, R_{\nu,6}^{-1/3}
L_{\nu,51}^{1/6}\epsilon_{\nu,MeV}^{1/3} \left(y_\nu/y\right)^{1/3} $,
where $L_{\nu,51}$ and $\epsilon_{\nu,MeV}$ are the electron neutrino
luminosity and average neutrino energy at the neutrino sphere in units
of $10^{51} \, \textrm{ergs s}^{-1}$ and MeV, respectively.  The
gravitational redshift factor is $y = \sqrt{1-2GM_{NS}/r c^2}$ which, 
when evaluated at the neutrino sphere, $R_\nu$, is $y_\nu$.  Notice that
the only dependence on radius is carried in the redshift factor, so that
the atmosphere is close to isothermal.
 
At the radius, $r_c$, where the pressure in the envelope becomes
radiation dominated, the material becomes unstable to outflow
\citep{Salpeter81}.  Since the neutrino luminosity is significantly
lower than the neutrino Eddington luminosity, a thermally driven wind
results \citep{Duncan86}.  The density at which this wind begins can
be found approximately by equating the radiation pressure to the
baryonic pressure.  This results in a critical density, $ \rho_c
\approx \sci{8.3}{7} \, \textrm{g} \, \textrm{cm}^{-3} \,
R_{\nu,6}^{-1} L_{\nu,51}^{1/2}\epsilon_{\nu,MeV}
\left(y_\nu/y\right), $ at which significant outflow begins and the
kinetic equilibrium of weak interactions ceases to hold.  Under these
conditions, nuclear statistical equilibrium is maintained on a time
scale much shorter than the dynamical time scale and, for these
temperatures and densities, there will be no bound nuclei present.
Since the electron fraction is set by kinetic equilibrium, the
composition of the wind does not depend on any previous nuclear
processing, so any nucleosynthesis from the wind will be primary.

Assuming that most neutrino heating occurs near $r_c$, the entropy is
constant once the temperature cools to the nucleon recombination
temperature, $kT \approx 0.5$ MeV.  Therefore, the final nuclear
abundances in the wind depend mainly on the wind entropy, electron
fraction, and the dynamical timescale at the radius where alpha 
combination occurs \citep{Qian96,Hoffman97}.  To
determine the contribution of the wind to the nucleosynthesis of the
entire supernova, the mass loss rate must also be known.  Estimates
for these quantities are given in the Appendix along with a discussion
of the effect of general relativistic corrections.

Integrating the mass loss rate (equation \ref{eq:mdot}) for a typical
neutrino luminosity history implies that the wind will eject
approximately $10^{-3} \, M_\odot$ of material.  This in turn means
that for the wind to contribute to the integrated yields of the
supernova for a particular isotope, that isotope needs to overproduced
relative to its solar mass fraction by a factor of at least $10^5$ in
the wind, assuming the rest of the supernova ejects $\sim10 \, M_\odot$ and has
over production factors of its most abundant metals of order 10.

At early times in the wind, the PNS is losing lepton number
\citep[e.g.][]{Burrows86}.  Neutrino interactions in the wind then tend to
increase the lepton number of the wind, and, to maintain charge
neutrality, cause the wind to become proton rich.  Under these
conditions, the wind may synthesize some of the light p-process
elements via the so called $\nu$p-process \citep{Frohlich06,Pruet06}.

\begin{figure*}
\begin{center}
\includegraphics[scale=0.7]{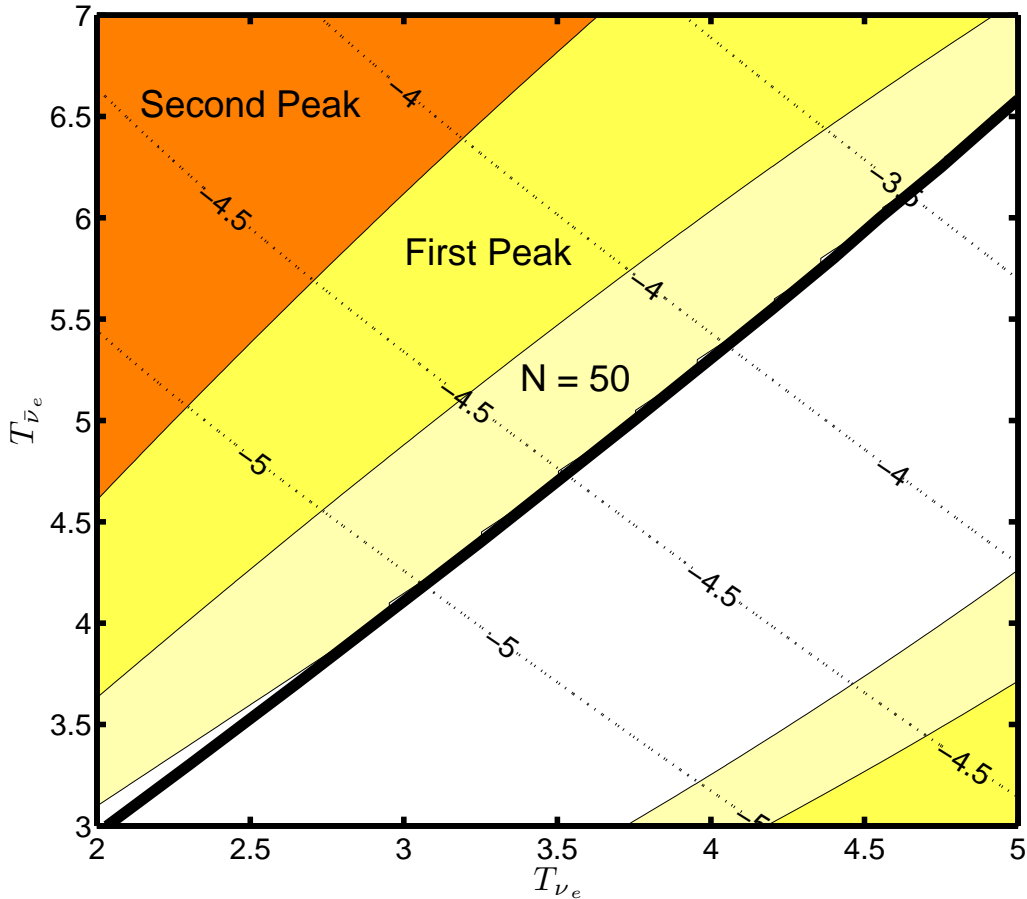}
\end{center}
\caption{Neutrino two-color plot produced using the analytic relations
  in the Appendix.  A neutron star with gravitational mass 1.4
  $M_\odot$ \ has been assumed with a neutrinosphere radius of 10 km.
  The total neutrino luminosity is assumed to scale as $L_{\nu_e,tot}
  = 10^{51} (\langle T_{\nu} \rangle/ 3.5 \, \textrm{MeV})^4 \,
  \textrm{erg} \textrm{s}^{-1}$.  This luminosity is split between
  neutrinos and anti-neutrinos so as to ensure that the net
  deleptonization rate of the PNS is zero.  The thick black line
  corresponds to an electron fraction of $Y_e = 0.5$.  Above this
  line, neutron-rich conditions obtain and below it the matter is
  proton-rich.  The white region is where there no free neutrons
  remain after charged particle reactions cease.  The N = 50 (tan)
  region corresponds to final neutron-to-seed ratios between 0.01 and
  15.  The ``first peak'' (yellow) region corresponds to a neutron-to-seed
  ratio between 15 and 70, and the ``second peak'' (orange) region is where
  the neutron-to-seed ratio is greater than 70.  The dashed lines
  correspond to the base ten logarithm of the mass loss rate in solar
  masses per second. }
\label{fig:tcp_base}
\end{figure*}

After the initial deleptonization burst though, the net lepton number
carried by the wind will be small. Since the anti-electron
neutrinosphere sits deeper in the PNS than the electron
neutrinosphere, the electron neutrinos will be cooler than the
electron anti-neutrinos \citep{Woosley94}.  If this asymmetry is large
enough, the wind can become neutron rich at later times.  Combined
with an $\alpha$-rich freeze out, this can give conditions favorable
for the r-process \citep{Woosley92}.

In both cases, the resulting nucleosynthesis is characterized by the
integrated neutron to seed ratio after charged particle reactions
freeze out.  For proton-rich winds, alpha-particles recombine into
$^{12}$C by the standard triple alpha reaction and then alpha-capture
and proton-capture reactions carry the nuclear flow up to
approximately mass 60 \citep{Woosley92}.  The slowest reaction in this
sequence is $^4$He($2\alpha$,$\gamma$)$^{12}$C, so the total number of
seed nuclei produced is equal to the number of $^{12}$C nuclei
produced.  When only free protons are present, the integrated neutron density is
determined by the rate of anti-neutrino capture on free protons.  An
estimate for the neutron to seed ratio for proton-rich winds is given
by equation \ref{eq:nsp}.  A neutron to seed ratio of only a few is
required to bypass the few long-lived waiting points which hinder
production of the light p-isotopes \citep{Pruet06}.  Still, it is
challenging to produce even a small neutron to seed ratio, since the
dynamical time scale of the wind is short compared to the anti-neutrino
capture time scale.

In the neutron-rich case, seed nuclei are produced by a different
reaction sequence $^4$He($\alpha$n,$\gamma$)$^9$Be($\alpha$,n)$^{12}$C
\citep{Woosley92}.  For the conditions encountered in the wind, the
neutron catalyzed triple-alpha reaction proceeds about ten times as
quickly as $^4$He($2\alpha$,$\gamma$)$^{12}$C.  This increases the
seed number compared with that obtained in a proton-rich wind with
similar dynamical properties.  Also, since there are free neutrons,
capture can proceed up to the N=50 closed shell isotopes $^{88}$Sr,
$^{89}$Y, and $^{90}$Zr.  Here, the neutron density is just determined
by the number of free neutrons left after charged particle reactions
freeze out.  The neutron to seed ratio in neutron-rich winds can be
approximated using equation \ref{eq:nsn}.  Charged particle reactions
continue up to the N=50 closed shell, at which point it becomes
unfavorable to capture alpha particles due to small separation
energies and large coulomb barriers \citep{Hoffman96}.  If neutrons are exhausted during
this process, the wind will mainly produce the isotopes $^{88}$Sr,
$^{89}$Y, and $^{90}$Zr \citep{Hoffman97}.  This happens when the
condition
\begin{equation} 
\label{eq:n50_prdod} 
\frac{\bar Z}{\bar A} \approx 0.42-0.49 = \frac{Y_e
  f_\alpha}{2Y_e(f_\alpha-1) + 1}
\end{equation} 
is met.  Here, $f_\alpha \approx 14 Y_s/Y_{\alpha,i}$ is the fraction
of the initial helium abundance that gets processed into heavy nuclei.
Neutron to seed ratios of approximately 30 and 110 are required to
produce first and second peak r-process nucleosynthesis, respectively.

Using the analytic results for the wind dynamics and nucleosynthesis
given in the Appendix (equations \ref{eq:rcrit}, \ref{eq:ent},
\ref{eq:mdot}, \ref{eq:tau},\ref{eq:ye}, \ref{eq:nsp}, \ref{eq:nsn},
and using the neutrino interaction rates given in
\S\ref{sec:neutrino_rates} to fix the thermodynamic state at $r_c$),
one can easily explore the neutrino temperature parameter space to
determine the neutrino temperatures and fluxes that are most conducive
to the r-process or the production of the light p-process.  Figure
\ref{fig:tcp_base} is a neutrino two-color plot where it is assumed
that the deleptonization rate is zero and that the neutrino luminosity
scales with the temperature to the fourth power ($L_{\nu_e,tot} =
10^{51} (\langle T_{\nu} \rangle/ 3.5 \, \textrm{MeV})^4 \,
\textrm{erg} \textrm{s}^{-1}$).  The different nucleosynthetic regions
are delineated by the final calculated neutron to seed ratio.  To give
a feeling for how a particular point in parameter space might
contribute to the integrated nucleosynthesis of the wind, the mass
loss rate is also shown.

For a significant amount of material to move past the N = 50 closed
shell during neutron-rich conditions, the anti-neutrino temperature
must be approximately $60$\% higher than the neutrino temperature.
For second peak r-process nucleosynthesis to occur, the asymmetry must
be greater than $100$\%.  Modern PNS cooling calculations do not give
such large asymmetries \citep{Pons99,Huedepohl09}.

Under proton-rich conditions, only a small region of the
parameter space at high neutrino and low anti-neutrino temperature is
favorable for the $\nu$p-process.  There will be a small amount of
neutron production in the white region, but it is unlikely that
significant production of the light p-process elements $^{74}$Se,
$^{78}$Kr, $^{84}$Sr, and $^{92}$Mo will occur.  The region in
neutrino temperature space where there is significant neutron
production is unlikely to be reached.  This region is small due to the
short dynamical time scale of the wind, which reduces the time over
which anti-neutrinos can capture on free neutrons.  One should note
that, very soon after shock formation in the supernova, a wind
solution may not be appropriate and material will be entrained closer
to the PNS for a longer period of time.  This scenario would be similar
to the the conditions used in \cite{Pruet06}.
  
Therefore, based upon simple principles, it seems unlikely that the
standard wind scenario will produce r-process or light p-process
isotopes in solar ratios, as is required by observations of metal poor
halo stars \citep{Sneden96}.  This same conclusion has been reached by
other authors \citep{Hoffman97,Thompson01}, but is repeated
here in simple terms.  We will find that our numerical calculations
give similar results and that there is no significant r-process
nucleosynthesis associated with the wind.  Still, the wind can produce 
some isotopes that may have an observable signature.  For standard PNS 
luminosities, the wind will spend a significant amount of time in the 
region of parameter space were N = 50 closed shell nucleosynthesis occurs.


\section{Computational Method}
\label{computational_method}

To more accurately investigate the integrated nucleosynthesis of the
NDW, we have updated the implicit Lagrangian hydrodynamics code Kepler
\citep{Weaver78,Woosley02} to carry out time-dependent simulations of
the wind dynamics and nucleosynthesis.  Kepler has been used
previously to study time-independent winds\citep{Qian96}, but the weak
and nuclear physics employed there was rudimentary and nucleosynthesis
was not tracked.  Trajectories from Kepler were used for post-processing
calculations of nucleosynthesis in \cite{Hoffman97}.

Kepler solves the non-relativistic hydrodynamic equations in
Lagrangian coordinates assuming spherical symmetry.  First order
general relativistic corrections are included in the gravitational
force law (cf. \cite{Shapiro83}).  All order $v/c$ effects are
neglected.  This is justified since the maximum wind speeds
encountered are, at most, a few percent of the speed of light.  The
momentum equation is then 
\begin{equation} 
\td{v_r}{t} = -4 \pi r^2 \pd{P}{m} - \frac{Gm}{r^2}\left( 1 +
\frac{P}{\rho c^2} + \frac{4 \pi P r^3}{m c^2} \right) \left(1 -
\frac{2 G m}{r c^2} \right)^{-1}
\end{equation}
where the symbols have their standard meanings.  As has been shown by
previous studies \citep{Qian96,Cardall97,Otsuki00,Thompson01}, general
relativistic corrections to the gravitational force can have an
appreciable effect on the entropy and dynamical time scale of the
wind.  The equation of state includes a Boltzmann gas of nucleons and
nuclei, an arbitrarily relativistic and degenerate ideal electron gas,
and photons.

\subsection{Weak Interaction Physics}
\label{sec:neutrino_rates}

Energy deposition from electron neutrino capture on nucleons, neutrino
annihilation of all neutrino flavors, and neutrino scattering of 
all flavors on electrons is included in the
total neutrino heating rate.  Neutrino ``transport'' is calculated in the 
light-bulb approximation.  The energy deposition rate is dominated
by neutrino captures on nucleons.  The neutrino annihilation rates
given in \cite{Janka91} are employed.  For the scattering rates, the
rates given in \cite{Qian96} are used, but we include general
relativistic corrections.  Standard neutrino capture rates are
employed in the limit of infinitely heavy nucleons with first order
corrections.  In this limit, the cross section is 
(Y.Z. Qian, private communication)
\begin{equation}
  \sigma_{ \nu n \atop \bar \nu p} =\frac{G_F^2 cos^2(\theta_C)}{ \pi
    (\hbar c)^4}\left[g_V^2+3g_A^2 \right] \left(\epsilon_{\nu} \pm
  \Delta \right)^2\left(1 \pm W_{M, {\nu \atop {\bar \nu}}}
  \epsilon_{\nu}\right) 
\end{equation} 
Here, $G_F$ is the Fermi coupling constant, $\theta_C$ is the Cabibo angle,
$g_V$ and $g_A$ are the dimensionless vector and axial-vector coupling 
constants for nucleons, $\Delta$ is the proton neutron mass difference, 
$\epsilon_\nu$ is the neutrino energy, and $W_{M, {\nu \atop {\bar \nu}}}$ 
accounts for the weak magnetism
and recoil corrections to the neutrino-nucleon cross section when the
base cross section is derived in the limit of infinitely heavy
nucleons \citep{Horowitz02}.  This correction reduces the
anti-neutrino cross section and increases the neutrino cross section
(by about a total of 10\% at the energies encountered in NDWs), which,
for a given incident neutrino spectrum, significantly increases the
asymptotic electron fraction.  Assuming a thermal distribution, these
cross sections result in the neutrino energy deposition rate for
anti-electron neutrino capture 
\begin{equation} 
\begin{array}{rl}
    \dot q_{\bar \nu p}=&\sci{4.2}{18} \textrm{ergs
      s}^{-1}\textrm{g}^{-1}\frac{Y_p L_{\bar \nu,51}}{ \langle
      \mu\rangle r_6^2} \\ \times & \biggl[-W_M^{\bar\nu
        p}\frac{\langle \epsilon_{\bar \nu}^4 \rangle }{\langle
        \epsilon_{\bar \nu} \rangle } + (1 + 2W_M^{\bar\nu p} \Delta)
      \frac{\langle \epsilon_{\bar \nu}^3 \rangle }{\langle \epsilon_{\bar\nu}
        \rangle } \\ & - (2 \Delta + W_M^{\bar\nu p} \Delta^2
      )\frac{\langle \epsilon_{\bar \nu}^2 \rangle }{\langle \epsilon_{\bar \nu}
        \rangle } + \Delta^2 \biggr]
\end{array}
\end{equation} 
and a similar expression for electron neutrino capture.  The neutrino
energy distributions are parameterized by assuming a Fermi-Dirac
spectrum.  The neutrino energy averages, $\avg{\epsilon_\nu^n}$, are
evaluated using this distribution.  The neutrino energy moments and
luminosity are evaluated in the rest frame of the fluid.  With general
relativistic corrections for the bending of null geodesics, the
average neutrino angle is given by
\begin{equation} 
\langle \mu \rangle = \frac{1}{2} +
\frac{1}{2}\sqrt{1-\left( \frac{R_\nu y_\nu}{r y}\right)^2}.
\end{equation} 
Special relativistic corrections are negligible in the regions where
neutrino interactions are important.

The lepton capture rates used are calculated in the limit of
infinitely heavy nucleons.  This results in a positron capture energy
loss rate
\begin{equation}
\begin{array}{rl}
\dot q_{e^+ n} =&\sci{6.9}{15}\, \textrm{ergs g}^{-1} \, \textrm{s}^{-1} \, Y_n T_{10}^6 \\
\times & \int_0^\infty du f_{e}(u,-\eta)  \left( u^5 + 3 \delta u^4 + 3 \delta^2 u^3 + \delta^3 u^2  \right)
 \end{array}
\end{equation}
here $f_e(u,\eta) = (\exp(u-\eta)+1)^{-1}$, $\eta$ is the electron degeneracy parameter, 
$\delta$ is the proton neutron mass difference divided by $k_bT$, and $Y_n$ is the neutron 
fraction.  A similar rate is employed for electron capture.

For the neutrino losses, we include electron and positron capture on
nucleons and include thermal losses as tabulated in \cite{Itoh96}.
The energy loss rate in the wind is dominated by the electron
captures.

\subsection{Nuclear Physics}

During a hydrodynamic time step in Kepler, the nuclear energy generation
rate and the changing nuclear composition are calculated
using a modified version of the 19-isotope network described in
\cite{Weaver78}.  Neutrino and electron capture rates on nucleons are
coupled to the network, which are calculated under the same
assumptions as the charged current energy deposition/loss rates
described above.  Therefore, non-equilibrium evolution of the electron
fraction is accurately tracked.

Although this network is appropriate for calculating energy generation
throughout the entire wind, it is not large enough to accurately track
the nucleosynthesis once alpha recombination begins at $ T \approx
0.5$ MeV.  Therefore, for temperatures below $20 \textrm{GK}$ an
adaptive network is run alongside the hydrodynamics calculation.  The
details of this network can be found in \cite{Woosley04} and 
\cite{Rauscher02}.  As a fluid
element passes the temperature threshold, the composition from the
19-isotope network is mapped into the adaptive network.  Typically,
the network contains approximately 2000 isotopes.  Where available,
experimental nuclear reaction rates are employed, but the vast
majority of the rates employed in the network come from the
statistical model calculations of 
\cite{Rauscher00}.  In general, the nuclear physics employed in these
calculations is the same as that used in \cite{Rauscher02}.  The
nucleon weak interaction rates employed in the 19-isotope network are
also used in the adaptive network.

\subsection{Problem Setup and Boundary Conditions}

To start the neutrino driven wind problem, an atmosphere of mass 0.01
$M_\odot$ is allowed to relax to hydrostatic equilibrium on top of a
fixed inner boundary at the neutron stars radius.  The mass enclosed
by the inner boundary is the neutron star's mass.  The photon
luminosity from the neutron star is assumed to be nearly Eddington,
but we have found that the properties of the wind are insensitive to
the the luminosity boundary condition.  Once hydrostatic equilibrium
is achieved, the neutrino flux is turned on and a thermal wind forms.
This wind is allowed to relax to a quasi-steady state, and then the 19
isotope network is turned on and the wind is, once again, allowed to
reach a quasi-steady state.  After this point, the neutrino flux is
allowed to vary with time, and the adaptive network is turned on.  

As the calculation proceeds, the mass of the envelope being followed
decreases and could eventually all be blown away.  To prevent this,
mass is added back to the innermost mass elements at a rate equal to
the mass loss rate in the wind.  The mass added to a fluid element at
each time step is a small fraction of its total mass.  We find that
mass recycling has no effect on the properties of the wind. It is
simply a way of treating a problem that is essentially Eulerian in a
Lagrangian code.

For most runs, a zero outer boundary pressure and temperature are 
assumed.  To investigate the effect of a wind termination shock, a
time dependent outer boundary condition is included in some of 
the simulations detailed below.  The pressure 
of the radiation dominated region behind the supernova 
shock is approximately given by \citep{Woosley02}
\begin{equation}
\label{eq:Pbound}
P_{ps} \approx \frac{E_{sn}}{4 \pi (v_{sn}  t)^3}
\end{equation}   
where $E_{sn}$ is the explosion energy of the supernova, $v_{sn}$ is
the supernova shock velocity, and $t$ is the time elapsed since the 
shock was launched.  As was discussed in \cite{Arcones07}, this 
results in a wind termination shock at a radius where the condition
$\rho_w v_w^2 + P_w \approx P_{ps}$
obtains, where $v_w$ is the wind velocity and $\rho_w$ is the 
wind density.  To avoid an accumulation of too many zones, 
mass elements are removed from the calculation once they 
exceed a radius of $10,000 \, \textrm{km}$.  This is well 
outside the sonic point and nuclear burning has 
ceased by this radius in all calculations .
 
\section{Numerical Results}
\label{results}

To survey both low and intermediate mass core collapse supernovae, 
neutrino emission
histories were taken from two core collapse calculations, one from a
$20 M_\odot$ \citep{Woosley94} supernova calculation and the other 
from a $8.8 M_\odot$ \citep{Huedepohl09}
supernova calculation.  Since the PNSs studied have significantly
different masses and neutrino emission characteristics, one is able to
get a rough picture of how integrated nucleosynthesis in the NDW
varies with progenitor mass.

\subsection{Neutrino Driven Wind from a $20 M_\odot$ Supernova}

\begin{figure}
\begin{center}
\leavevmode
\includegraphics[scale=0.45]  {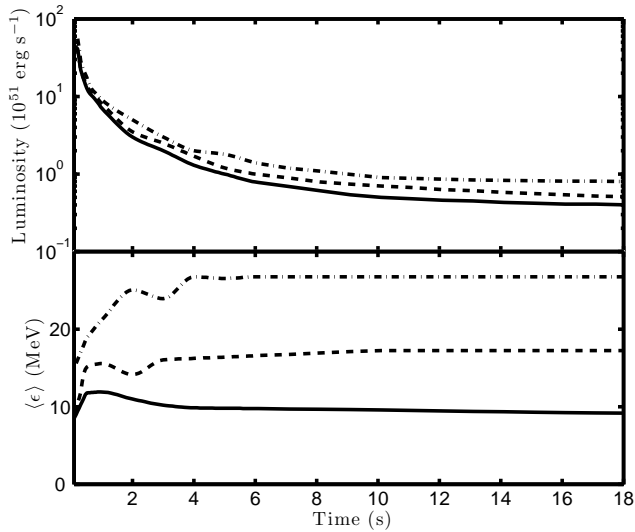}
\end{center}
 \caption{Neutrino luminosities and temperatures taken from the model
  of \cite{Woosley94}.  The top panel is the neutrino luminosities.  The 
  bottom panel is the average neutrino energies.  The solid line corresponds 
  to $\nu_e$, the dashed line corresponds to $\bar \nu_e$, the dot-dashed 
  line corresponds to $\nu_{\mu,\tau}$.}
 \label{fig:woos94_neut}
\end{figure}

The first set of neutrino luminosities and temperatures are taken from
\cite{Woosley94}.  This calculation began with a $20 M_\odot$ progenitor
meant to model the progenitor of 1987A \citep{Woosley88}.  The
resulting neutron star had a gravitational mass of $1.4 M_\odot$ and
the neutrino sphere was taken to be at 10 km.  The neutrino
luminosities and average energies as a function of time from this
model are shown in figure \ref{fig:woos94_neut}.  After about 4
seconds, the neutrino energies become constant and the large
difference between the electron neutrino and anti-neutrino energies
implies that the wind will be neutron rich.  This supernova model had
some numerical deficiencies (Sam Dalhed, Private Communication). The
entropy calculated for the wind in \cite{Woosley94} ($S/N_Ak \approx 400$)
were unrealistically large due to some problems with the equation of
state. Here, that is not so important because the NDW is being
calculated separately, but this study did rely on older neutrino
interaction rates and did not include weak magnetism corrections (see \S
\ref{sec:neutrino_rates}).  Therefore, the results obtained using
these neutrino histories are only suggestive of what might happen in a
more massive star.  If weak magnetism were taken into
account, the calculated electron and anti-electron neutrino
temperatures would probably be somewhat further apart.

\begin{figure}
\begin{center}
\leavevmode
\includegraphics[scale=0.5]  {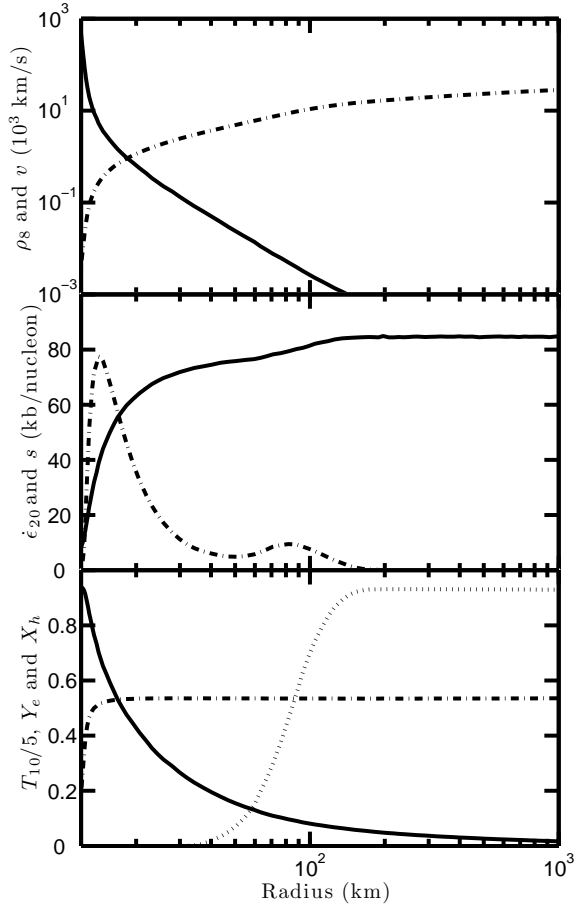}
\end{center}
\caption{Wind structure after two seconds in the model using 
the neutrino luminosities from \cite{Woosley94}.  The top panel 
shows the density in units of $10^8 \, \textrm{g} \, \textrm{cm}^{-3}$ 
(solid line) and the radial velocity in units of $10^3 \, \textrm{km} \, 
\textrm{s}^{-1}$ (dot-dashed line).  The middle panel shows the net 
energy deposition rate from weak and strong interactions in units of 
$10^{20} \, \textrm{erg} \, \textrm{g}^{-1} \, \textrm{s}^{-1}$ (dot-dashed line) 
and the entropy (solid line).  The bottom panel shows the temperature 
in units of $\sci{2}{9} \, \textrm{K}$ (solid line), the electron 
fraction (dot-dahsed line), and the fraction of material contained in 
nuclei (dotted line). }
 \label{fig:wnd_struct}
\end{figure}

The calculation was run for a total of 18 seconds.  During this time,
the mass loss rate decreased by almost three orders of magnitude while
a total mass of $\sci{2}{-3} M_\odot$ was lost in the wind.  A
snapshot of the wind structure two seconds after bounce is shown in
figure \ref{fig:wnd_struct}.  Note that the wind velocity stays very
sub-luminal throughout the calculation.  Therefore, the neglect of
special relativistic effects is reasonable.  The secondary bump in the
energy deposition rate occurs at the same radius where nucleons and
alpha-particles assemble into heavy nuclei.  This increases the
entropy by about 10 units.  Clearly, the electron fraction is set
interior to were nuclei form.  The radius where nuclei form is at a
large enough value that the alpha effect \citep{Fuller95} is not
significant at early times in the wind.  However, as the neutrino luminosity
decreases with time, nucleon recombination occurs at a smaller radius,
and the alpha effect becomes increasingly important.

\begin{figure}
\begin{center}
\leavevmode
\includegraphics[scale=0.43]  {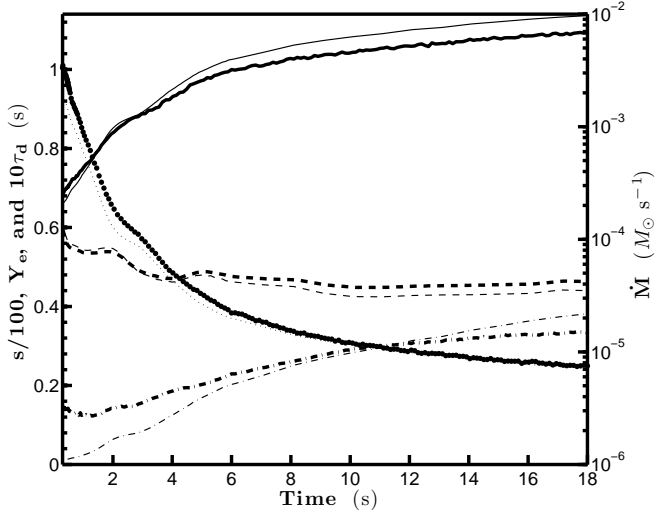}
\end{center}
\caption{Properties of the neutrino driven wind from the 
\cite{Woosley94} supernova model as a function of time. The thick 
lines correspond to the numerical results from Kepler and the thin 
lines correspond to the predictions of the analytic estimates described 
in the appendix.  The solid line is the dimensionless entropy per baryon, 
the dashed line is the electron fraction, the dash dotted line is the 
dynamical timescale, and the dotted line is the mass loss rate.  All 
of the quantities are taken extracted from where the wind temperature 
reaches 2 GK.}
 \label{fig:woos94_wndprop}
\end{figure}

The time evolution of the wind as calculated by Kepler is shown in
figure \ref{fig:woos94_wndprop}.  The increase in asymptotic entropy is mainly
driven by the decrease in neutrino luminosity, since the average
neutrino energies do not vary greatly.  The analytic approximation
(calculated using equation \ref{eq:ent} and the neutrino interaction
rates given in \S\ref{sec:neutrino_rates}) to the entropy tracks the
entropy calculated in Kepler fairly well.  This implies that the
variation in the neutrino luminosity with time does not significantly alter the
dynamics from a steady state wind. In contrast to the high entropies
reported in \citet{Woosley94}, the entropy here never exceeds 130.
For the time scales and electron fractions also obtained, such a low
value of entropy is not sufficient to give a strong r-process (see
below).

The electron neutrino and anti-neutrino energies do move further apart
as a function of time though, which causes the wind to evolve from
proton-rich conditions at early times to neutron-rich conditions
later.  A transition occurs from the synthesis of proton-rich isotopes
via the $\nu p$-process at early times to the $\alpha-$process mediated
by the reaction sequence
$\alpha$($\alpha$n,$\gamma$)$^9$Be($\alpha$,n)$^{12}$C later.  The
slight difference between the analytic approximation and the Kepler
calculation of $Y_e$ is due to the alpha effect \citep{Fuller95}.

\begin{figure}
\begin{center}
\leavevmode
\includegraphics[scale=0.65]  {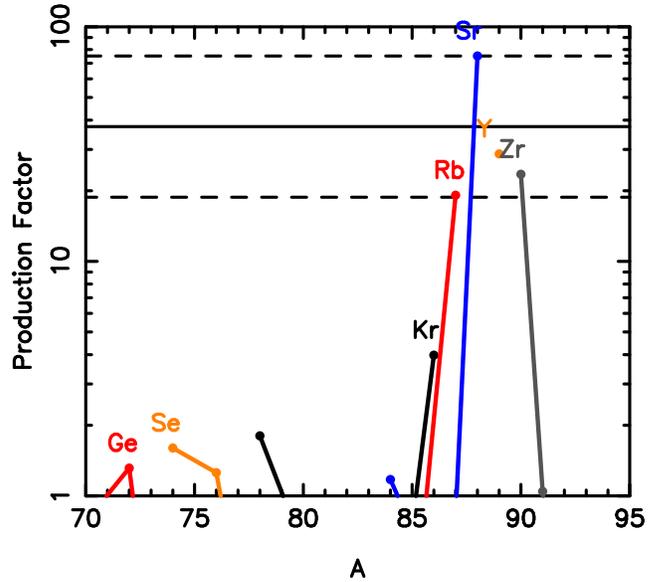}
\end{center}
\caption{Isotopic production factors from the NDW model when  
the neutrino luminosities from \cite{Woosley94} are used.  The production 
factors are calculated assuming that $18.4 \, M_\odot$ of material 
was ejected in the supernova in addition to the wind.  The top dashed line
corresponds to the greatest production factor in the wind, the solid line is a
factor of two below that, and the bottom dashed line is a factor of two below the
solid line.  These lines specify an approximate coproduction band for the wind alone.}
 \label{fig:WWsolopf}
\end{figure}

Integrated production factors for the wind are shown in figure
\ref{fig:WWsolopf}.  The production factor for the species $i$ is
defined as
\begin{equation} 
P_i = \frac{X_{i,w}M_{w}}{X_{i,\odot}(M_{w}+M_{sn})},
\end{equation} 
where $X_{i,w}$ is the mass fraction of species $i$  in the wind
after all material has decayed to stable isotopes, $M_w$
is the mass ejected in the wind, and $M_{sn}$ is the amount of mass
ejected by the entire supernova.  $X_{i,\odot}$ is the mass fraction
of isotope $i$ in the sun for which the values of \citet{Lodders03}
were used. The only isotopes that are co-produced in the wind
alone are $^{87}$Rb, $^{88}$Sr, $^{89}$Y, and $^{90}$Zr, with
production factor of $^{88}$Sr about a factor of 3 higher than the
other two N = 50 closed shell isotopes.  Before eight seconds, the
production factors had been much closer.  After eight seconds though,
the wind is dominated by $^{88}$Sr because $Y_e \sim 0.45$ and only
53\% of alpha particles are free after freeze out which puts
$\frac{\bar Z}{\bar A}\approx 0.41 $ of the heavy nuclei just
below the range given in equation \ref{eq:n50_prdod}.  There are not
enough free neutrons to make any significant amount of heavier nuclei,
and this results in significant production of the stable N = 50 closed
shell isotope with the lowest $\frac{\bar Z}{\bar A}$.

\begin{figure}
\begin{center}
\leavevmode
\includegraphics[scale=0.42]  {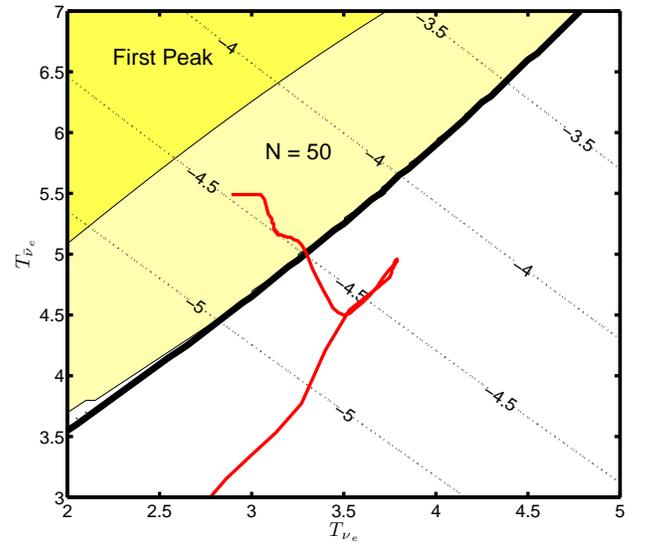}
\end{center}
 \caption{Neutrino two-color plot when the anti-neutrino luminosity 
 is 1.2 times neutrino luminosity, and the total luminosity scales 
 with average temperature to the fourth.  Similar to figure 
 \ref{fig:tcp_base}.  The red line is the neutrino temperatures 
 from \cite{Woosley94}.}
 \label{fig:tcp_woos}
\end{figure}

\begin{figure}
\begin{center}
\leavevmode
\includegraphics[scale=0.65]  {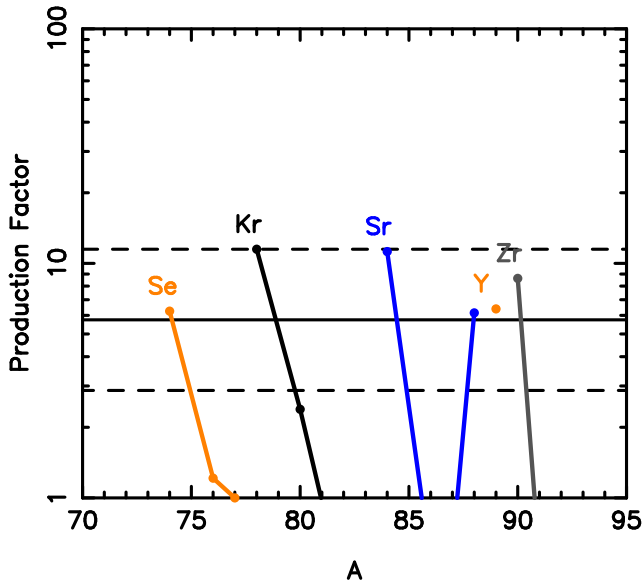}
\end{center}
\caption{Isotopic production factors from the NDW model 
employing the neutrino luminosities from \cite{Woosley94} with 
the anti-electron neutrino temperature reduced by $15\%$.  The 
production factors are calculated assuming that $18.4 \, M_\odot$ 
of material was ejected in the supernova in addition to the wind.
The horizontal lines are similar to those in figure \ref{fig:WWsolopf}.}
 \label{fig:WWredpf}
\end{figure}

\begin{figure}
\begin{center}
\leavevmode
\includegraphics[scale=0.65]  {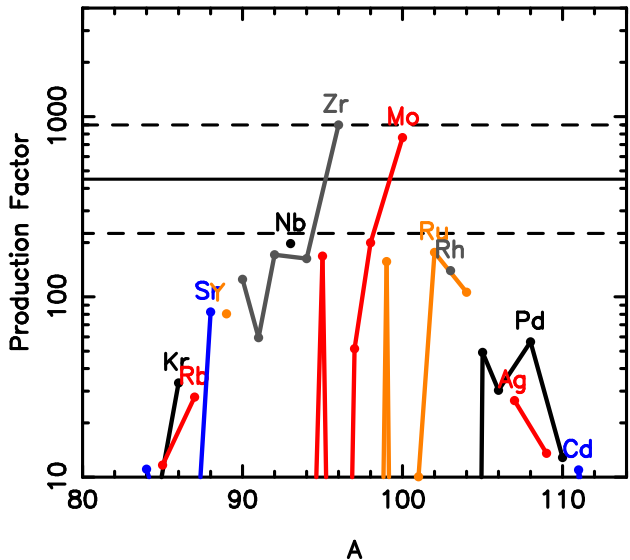}
\end{center}
\caption{Isotopic production factors from the NDW model 
employing the neutrino luminosities from \cite{Woosley94} 
with weak magnetism corrections turned off.  The production 
factors are calculated assuming that $18.4 \, M_\odot$ of material 
was ejected in the supernova in addition to the wind.
The horizontal lines are similar to those in figure \ref{fig:WWsolopf}.}
 \label{fig:WWnowmpf}
\end{figure}

\begin{figure}
\begin{center}
\leavevmode
\includegraphics[scale=0.65]  {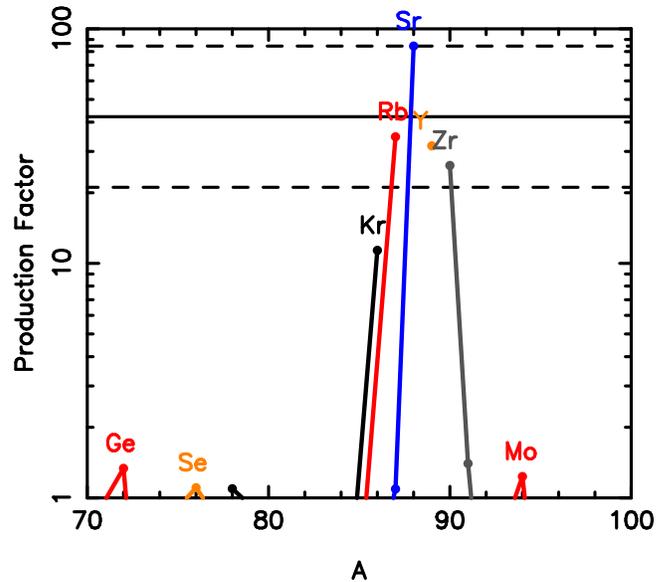}
\end{center}
\caption{Isotopic production factors from the NDW model when  
the neutrino luminosities from \cite{Woosley94} are used and an external boundary 
pressure is specified as described in the text, which results in a wind termination
shock.  The production 
factors are calculated assuming that $18.4 \, M_\odot$ of material 
was ejected in the supernova in addition to the wind.  The horizontal lines are 
similar to those in figure \ref{fig:WWsolopf}.}
 \label{fig:WWpbd}
\end{figure}

During the first four seconds, the wind is proton rich and the isotopes
$^{69}$Ga, $^{70,72}$Ge, $^{74,76}$Se, and $^{78,80,82}$Kr 
are produced by proton captures on seed nuclei produced by the triple-alpha 
reaction and subsequent ($\alpha$,p) reactions.  Although the mass 
loss rate is much higher when the wind 
is proton rich, the alpha-fraction freezes out at 98\% of its initial value, 
which results in significantly decreased production of heavy nuclei.  
The difference in final alpha fraction between the neutron- and proton-rich 
phases of the wind is due mainly to the difference in speed of the reaction 
chains $\alpha$($2\alpha$,$\gamma$)$^{12}$C and
$\alpha$($\alpha$n,$\gamma$)$^9$Be($\alpha$,n)$^{12}$C, but also 
to the decreased entropy at early times.

We can compare this with the analytic predictions for nucleosynthesis
by plotting the neutrino temperature evolution from this model on a
neutrino ``two-color plot'' (figure \ref{fig:tcp_woos}).  Here we have
set $L_{\bar \nu_e} = 1.2 L_\nu$ which is approximately correct at
late times in the calculation of \cite{Woosley94}.  The wind never
reaches a region in which r-process nucleosynthesis is expected, but
spends a significant amount of time making nuclei in the N = 50 closed
shell isotones.

\subsubsection{Variations in Neutrino Properties}
Since the neutrino temperatures from the original model were
uncertain, several other models were calculated. One had a reduced (by $15\%$)
electron antineutrino temperature; another had the weak
magnetism corrections to the neutrino interaction rates turned off.  A
smaller antineutrino temperature is more in line with recent
calculations of PNS cooling \citep{Pons99,Keil03}.  Because the
model of \cite{Woosley94} did not include weak magnetism corrections,
our model with weak magnetism corrections turned off is more
consistent with the original supernova model.

The production factors for the model with a reduced electron
antineutrino temperature are shown in figure \ref{fig:WWredpf}.  The
yield of $^{88}$Sr is reduced by almost a factor of ten from the base
case, while the production factors of $^{89}$Y and $^{90}$Zr are
reduced by a factor of three. In this case, the wind also produces the
proton-rich isotopes $^{74}$Se, $^{78}$Kr, and $^{84}$Sr.  The
coproduction line for lighter elements like oxygen in a $20 M_\odot$
supernova at solar metallicity is around 18, so the wind could contribute
to the total nucleosynthesis if the antineutrino temperature was
reduced, but its contribution would be small.  

The yields when weak magnetism corrections are ignored are shown in
figure \ref{fig:WWnowmpf}.  Without weak magnetism, the electron
fraction drops below 0.4 at late times when the entropy is fairly
high.  Equation \ref{eq:n50_prdod} is no longer satisfied and material
moves past the N = 50 closed shell towards A $\approx$ 110.  Some
r-process isotopes are produced, such as $^{96}$Zr and $^{100}$Mo, but
not anywhere near solar ratios, and no material reaches the first
r-process peak.

\begin{figure*}
\begin{center}
\leavevmode
\includegraphics[scale=0.75]  {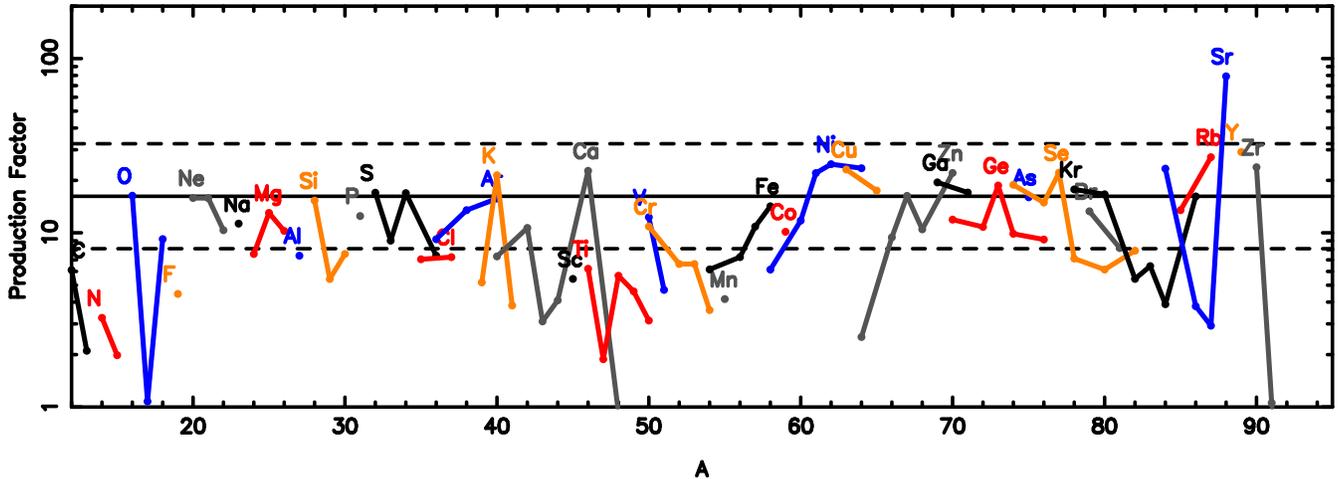}
\end{center}
\caption{Combined isotopic production factors of the neutrino 
driven wind with unaltered neutrino temperatures and including weak magnetism
corrections added to those of a $20 M_\odot$ stellar model from \cite{Woosley95}.  
The solid black line is the coproduction line with $^{16}$O.  
The dashed lines are a factor of two above and below the coproduction line.  The 
neutrino driven wind is responsible for the production of $^{88}$Sr, 
$^{89}$Y, and $^{90}$Zr.}
 \label{fig:comb_pf}
\end{figure*}

\subsubsection{Effect of a Wind Termination Shock}
To investigate the possible effect of a wind termination shock on nucleosynthesis,
another model was run with a boundary pressure and temperature 
determined by equation \ref{eq:Pbound}.  An explosion energy of 
$10^{51} \, \textrm{erg}$ was assumed and the shock velocity was taken as
$\sci{2}{9} \, \textrm{cm s}^{-1}$.  This resulted in a wind termination
shock that was always at a radius greater than $10^3 \, \textrm{km}$.  
The production factors from this model are shown in figure 
\ref{fig:WWpbd}.  Similar to the simulation without a wind termination
shock, the N=50 closed shell elements dominate the wind's nucleosynthesis.

The main difference between the case with and without a wind termination 
shock is a shift in the mass of isotopes produced during the proton-rich
phase.  This can be seen in the increased production of Mo. 
During this phase, the post shock temperature varied from 
2.5 GK down to 0.8 GK and the density varied from
$\sci{5}{4} \, \textrm{g cm}^{-3}$ to $\sci{5}{2} \, \textrm{g cm}^{-3}$.  
These conditions are very favorable for 
continued proton capture once the long lived waiting point isotopes 
$^{56}$Ni and $^{64}$Ge are bypassed by (n,p) reactions.  Because 
these conditions persist for at least a second after a fluid element passes
through the wind termination shock, significantly more proton captures 
can occur on seed nuclei that have moved past mass $\sim 64$ relative
to the case with no termination shock.  Still, not many more neutrons 
are produced per seed nucleus relative to the base run.  Therefore, the net 
number of seeds that get past the long lived waiting points remains small
and the proton-rich wind does not contribute to the integrated nucleosynthesis.     
It should also be noted that a different treatment of the wind's 
interaction with the supernova shock might result in a breeze solution 
which may supply more favorable conditions for $\nu$p-process 
nucleosynthesis \citep{Wanajo06}.    

\subsubsection{Total Supernova Yields}

In figure \ref{fig:comb_pf}, the production factors from a $20
M_\odot$ supernova model from \cite{Woosley95} have been combined with
the production factors we calculated in the NDW with the unaltered
neutrino histories of \citep{Woosley94} with weak magnetism corrections
included.  The wind could be
responsible for synthesizing the isotopes $^{87}$Rb, $^{88}$Sr,
$^{89}$Y, and $^{90}$Zr.  $^{88}$Sr production is above the
co-production band, but the rest are in agreement with the stellar
yields.  This overproduction of $^{88}$Sr is similar to the result of
\cite{Hoffman97}.

For the model with a reduced anti-electron neutrino temperature
combined with the yields from the $20 M_\odot$ supernova model, the
wind contributes 28\%, 42\%, 35\%, 75\%, 75\%, and 80\% of the total
$^{74}$Se, $^{78}$Kr, $^{84}$Sr, $^{88}$Sr, $^{89}$Y, and $^{90}$Zr
abundances in the supernova, respectively.  This wind model does not result in any
isotopes being overproduced relative to the rest of the yields of the
supernova.  For the case with weak magnetism turned off, the nuclei
produced by the wind are overproduced relative to those made in the rest of 
the star by factor of nearly 100, hence this
would need to be a very rare event if this model were realistic.

Clearly, weak magnetism corrections and variations in the neutrino
temperatures have a very significant effect on nucleosynthesis in the
wind.  Aside from the effects of an extra source of energy
(\ref{modifications}), the neutrino spectra are the largest current
theoretical uncertainty in models of the NDW.

\subsection{Neutrino Driven Wind from a $8.8 M_\odot$ Supernova}
\label{eight_results}

\begin{figure}
\begin{center}
\leavevmode
\includegraphics[scale=0.45]  {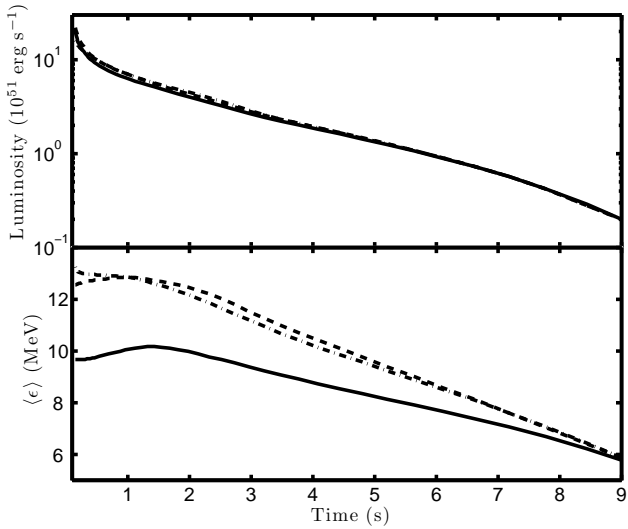}
\end{center}
 \caption{Neutrino luminosities and temperatures taken from the model 
 of \cite{Huedepohl09}.  The line styles are the same as in figure 
 \ref{fig:woos94_neut}.  }
 \label{fig:jnk_neut}
\end{figure}

The second PNS model is a more modern one-dimensional calculation of an
electron-capture supernova \citep{Huedepohl09} that started from an
$8.8 M_\odot$ progenitor model \citep{Nomoto84}. This resulted in a
PNS with a gravitational mass of $1.27 M_\odot$ and a radius of 15 km.
Together the lower mass and increased radius imply a lower
gravitational potential at the neutrinosphere.  This work employed
neutrino interaction rates which took weak magnetism and ``in-medium''
effects into account.  The neutrino luminosities and average energies
as a function of time are shown in figure \ref{fig:jnk_neut}. The
maximum difference between the electron and anti-electron neutrino
average energies is significantly less than in the model of
\cite{Woosley94}.  This is likely due in part to both the decreased
gravitational potential of the PNS and the more accurate neutrino
interaction rates in the newer model.

\begin{figure}
\begin{center}
\leavevmode
\includegraphics[scale=0.45]  {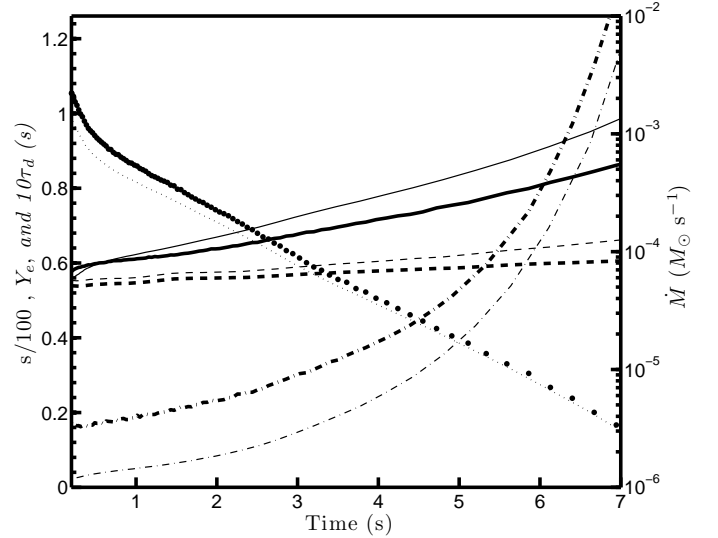}
\end{center}
\caption{Properties of the neutrino driven wind from the 
\cite{Huedepohl09} supernova model as a function of time.  
The lines have the same meaning as in figure \ref{fig:woos94_wndprop}.}
 \label{fig:jnk_wndprop}
\end{figure}

The calculation was run for a total of nine seconds, at which point
the mass loss rate had dropped by two orders of magnitude.  The total
amount of mass ejected in the wind was $\sci{3.8}{-4} \, M_\odot$.  In
figure \ref{fig:jnk_wndprop}, the properties of the NDW calculated
using Kepler are plotted as a function of time.  Notice that the
entropy never reaches above 100 in this model, which diminishes the
likelihood of significant nucleosynthesis.  For comparison, we also
include the analytic estimates detailed above.  There is reasonable
agreement between the analytic and the numerical calculations, but not
nearly as good as in the $20 M_\odot$ model.

In contrast to the simulation run with the neutrino luminosities of
\citet{Woosley94}, the electron fraction continues to increase with
time.  The difference between the average electron neutrino energy and
electron anti-neutrino energy is, at most, about 3 MeV, compared to a
maximum of 8 MeV in the \citet{Woosley94} calculations.  Also, the
difference between the average neutrino energies decreases as a
function of time, compared to an increase with time in
\citet{Woosley94}.  Finally, the energies of all kinds of neutrinos
are lower in the \citet{Huedepohl09} calculation, so that the
proton-neutron rest mass difference significantly suppresses the
anti-neutrino capture rate relative to the neutrino capture rate.
These differences are presumably due to both the different neutron
star masses and neutrino interaction rates employed.

\begin{figure}
\begin{center}
\leavevmode
\includegraphics[scale=0.65]  {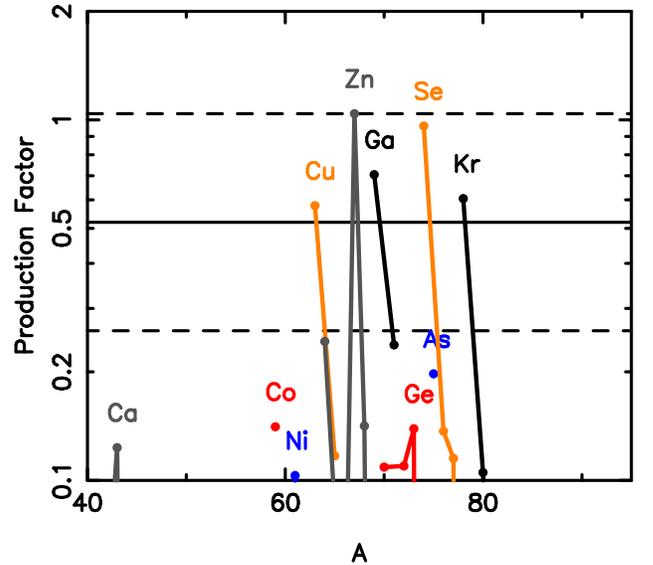}
\end{center}
\caption{Isotopic production factors from the NDW model employing 
the neutrino luminosities from \cite{Huedepohl09}.  The production 
factors are calculated assuming that $7.4 \, M_\odot$ of material was 
ejected in the supernova in addition to the wind.  
The horizontal lines are similar to those in figure \ref{fig:WWsolopf}.
Notice that none of the production factors are significantly greater 
than one.}
 \label{fig:jnkpf}
\end{figure}

The conditions in this model thus preclude {\sl any} r-process
nucleosynthesis, but they are potentially favorable for production of
some low mass p-process isotopes by the $\nu$p-process.  The
integrated isotopic production factors are shown in figure
\ref{fig:jnkpf}.  The total ejected mass was take as 7.4 $M_\odot$, as
1.4 $M_\odot$ neutron star is left behind in the calculation of
\cite{Huedepohl09}.  During the calculation a maximum network size of
988 isotopes is reached.  The p-process elements $^{74}$Se and
$^{78}$Kr are co-produced with $^{63}$Cu, $^{67}$Zn, and $^{69}$Ga,
but the maximum production factor for any isotope is 1 when weighted
with the total mass ejected in the supernova.  Therefore, in this
simple model, the proton-rich wind from low mass neutron stars 
will not contribute significantly to galactic chemical evolution.

The entropies encountered when the mass loss rate is high are low
($\sim 50$), so that there is more production of $^{56}$Ni by
triple-alpha and a subsequent $\alpha$p-process.  As the neutron
abundance available for the $\nu$p-process is given by
\begin{equation} 
Y_n \approx \frac{\lambda_\nu Y_p}{\rho N_A \sum_i
Y_i \avg{\sigma v}_{i(n,p)j}}, 
\end{equation} 
increased seed production reduces the available neutron abundance and
therefore hinders production of the p-process elements $^{74}$Se,
$^{78}$Kr, $^{84}$Sr, and $^{92}$Mo.  Additionally, at early times,
the dynamical time scale is short which implies a smaller integrated neutron 
to seed ratio, $\Delta_n$ (see the appendix).        

The yields of from this model cannot be combined with the yields from 
the rest of the supernova because they are not published.  As \cite{Nomoto84}
has discussed, the mass inside the helium burning shell was close to the 
mass of the neutron star that was left after the explosion.  Therefore 
the ejecta of the supernova is expected to have small production factors.
This implies that, even when the yields of the NDW are combined with the 
rest of the supernova, it is unlikely that these low mass core collapse
supernovae will contribute significantly to galactic chemical evolution.

\subsubsection{Effect of a Wind Termination Shock}

\begin{figure}
\begin{center}
\leavevmode
\includegraphics[scale=0.65]  {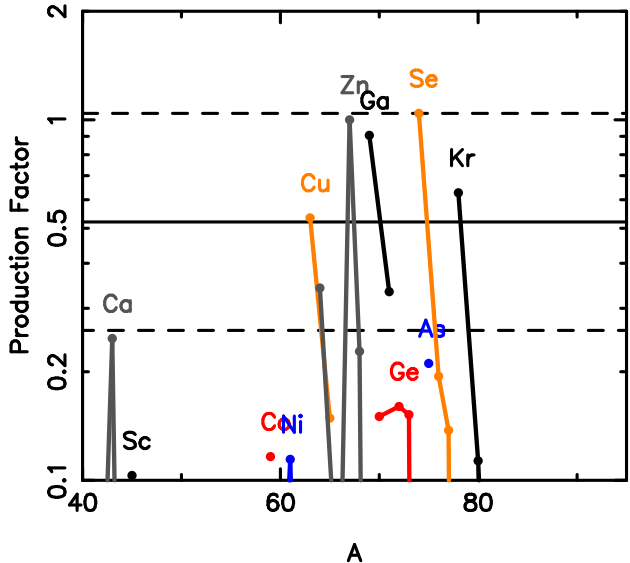}
\end{center}
\caption{Isotopic production factors from the NDW model employing 
the neutrino luminosities from \cite{Huedepohl09}, but including a time
dependent external boundary pressure which results in a wind 
termination shock.  The production 
factors are calculated assuming that $7.4 \, M_\odot$ of material was 
ejected in the supernova in addition to the wind.  
The horizontal lines are similar to those in figure \ref{fig:WWsolopf}.
Notice that the production factors are almost unchanged when an 
external boundary pressure is added.}
 \label{fig:jnkpfpbd}
\end{figure}

As was mentioned above, it is very possible that a transonic wind solution 
may not be appropriate this early in the supernovas evolution.  \cite{Fischer09} 
have found that a wind termination shock is not present in a one-dimensional
supernova model using the progenitor from \cite{Nomoto84}.  Still, it is 
interesting to consider the effect of a reverse shock on the wind nucleosynthesis.

A second simulation was run with a time dependent boundary pressure given
by equation \ref{eq:Pbound}, with $E_{sn} = 10^{50} \, \textrm{erg}$ and 
$v_{sn} = \sci{2}{9} \, \textrm{cm s}^{-1}$.  This results in a wind termination
shock at a radius of approximately $\sci{3}{8} \, \textrm{cm}$ throughout the 
simulation.  Inside the wind termination shock the wind dynamics are very similar
to those in the run with no boundary pressure.  The production factors from 
this model are shown in figure \ref{fig:jnkpfpbd}.  Clearly, there is almost no
difference in the nucleosynthesis in the runs with and without a wind termination
shock.
 
After $0.75 \, \textrm{s}$, the post shock temperature drops below 1 GK and 
the wind termination shock has little effect on subsequent nucleosynthesis.
Because the post shock temperature is high for less than one second, the 
wind termination shock has very little effect on the integrated nucleosynthesis.  
A larger explosion energy would likely result in a larger effect on the nucleosynthesis, 
but there are still very few neutrons available to bypass the long lived waiting points
and it seems unlikely that the production factors would be increased by 
more than a factor of a few.

\section{Discussion}
\label{discussion}

\subsection{Comparison with SN 1987A}

Since the progenitor model used in \cite{Woosley94} was a model for SN
1987A, it is interesting to compare our predicted abundances with
those observed in the ejecta of that event.  $^{88}$Sr, produced by
the NDW, dominates the elemental strontium yield when the results of
the 20 $M_\odot$ wind model are combined with those predicted by
\cite{Woosley88}, who ignored the wind.  For the base NDW model, [Sr/Fe]$=
0.8$, if weak magnetism corrections are neglected, [Sr/Fe]$= 1.6$; and
if the anti-neutrino temperature is reduced in the base model by 15\%,
[Sr/Fe]$=0.2$.

\cite{Mazzali92} found [Sr/Fe]$\approx 0.3$ in the ejecta of SN 1987A
2-3 weeks after the explosion. This observation has a substantial
error bar due to the uncertainty of modelling the spectrum of the
expanding ejecta. Of even greater concern is comparing of our models
for bulk yields with supernova photospheric abundances observed on a
given day when the observations are probably not even probing the
innermost ejecta.  Still, if one assumes that the [Sr/Fe] ratio found
by \cite{Mazzali92} represents the value for all of the ejecta (for
instance by assuming that ``mixing'' was extremely efficient), a weak
constraint can be put on the neutrino fluxes and energies predicted by
the model of \cite{Woosley94}. The NDW model with a reduced
anti-neutrino temperature is much closer to the observed value of
[Sr/Fe] than the other two models.  This suggests the anti-neutrino
temperature may have been overestimated in the original model, a
change that would be more consistent with more modern calculations of
neutrino spectra formation in PNS atmospheres \citep{Keil03}. But
obviously, a meaningful constraint will require a more complete modeling
of the multi-dimensional explosion and time-dependent spectrum of SN
1987A.

\subsection{Strontium and Yttrium in Halo Stars} 

Since strontium and yttrium are abundantly produced in our models, it
may be that the NDW has contributed to their production throughout
cosmic history. An interesting possibility is that the abundances of
these elements might trace the birth rate of neutron stars at an early
time.  Taking a standard r-process abundance pattern from metal poor
stars with strong r-process enhancments, \cite{Travaglio04} find that
8\% and 18\% of solar strontium and yttrium, respectively, are not
produced by either the ``standard'' r-process or any component of the
s-process.  It therefore seems plausible that charged particle
reactions in the NDW could make up this ``missing'' component.

Any nucleosynthesis that happens in the NDW will be primary,
i.e. provided that the mass function of neutron stars at birth does
not itself scale with metallicity, similar nucleosynthesis will occur
for stars of any population.  Below [Fe/H]$\sim -1.5$, no component of
the s-process contributes to the abundances of N = 50 closed shell
isotopes \citep{Serminato09}.  If the NDW escapes the potential well
of the PNS, and contributes to the galactic budget of N = 50 closed
shell isotopes, it should provide a floor to [Sr/Fe] and [Y/Fe].  Based
upon the arguments of \cite{Travaglio04}, this floor would be at
[Sr/Fe]$\approx-0.18$ and [Y/Fe]$\approx -0.16$.  These numbers assume
that when the main r-process source contributes in addition to the NDW,
[Sr/Fe] and [Y/Fe] approach their solar values even though the
s-process has yet to contribute.  This is consistent with
observations.

In defining this floor, one must assume that the abundances in a
particular star sample a large number of individual supernovae.
This is because the production of N=50 closed shell elements likely
depends on the PNS mass and therefore the progenitor mass.  As we have
found, [Sr/Fe]$=0.8$ in the $20 M_\odot$ model with reduced
anti-neutrino temperatures, but the $8.8 M_\odot$ model produces no
strontium.  Observations show that below [Fe/H]$\sim -3$, the spreads
in [Sr/Fe] and [Y/Fe] increase significantly and the mean values
falloff some \citep{Francois07,Cohen08,Lai08}.  Single stars have
values of [Sr/Fe] below the predicted floor.  This could be because, 
at this metallicity, the metals in a particular
star come from only a handful of supernovae.

Another possible explanation of this variation is that supernova fall
back varies with metallicity.  Since the NDW is the
innermost portion of the supernova ejecta, it will be the most
susceptible to fallback.  It has been found that the amount of
supernova fallback depends strongly on the metallicity of the
progenitor, especially going between zero and low metallicity
\citep{Zhang08}.  Additionally, mixing is also greatly reduced in zero
metallicity stars compared to solar metallicity stars due to the formers 
compact structure \citep{Joggerst09}. 

The current understanding of supernova fallback suggests that the
nucleosynthetic contribution of the NDW will be suppressed at very low
metallicity.  Of course, the ejection of iron by the supernova is also
very susceptible to fallback, so the effect of fallback on the
evolution of [Sr,Y/Fe] is complicated and may require fine tuning to
give the observed decrease.
A somewhat different explanation was offered by \citet{Qian08} who
attributed the fall off of [Sr/Fe] at low metallicity to the evolution
of the ``hypernova'' rate with metallicity. For their purposes,
hypernovae were stellar explosions that contributed iron without
making much strontium. 

Given the sensitivity of strontium and yttrium yields to uncertain NDW
characteristics, especially neutrino fluxes and temperatures, it may
be some time before the complex history of these elements is even
qualitatively understood. It is likely though that their abundances in
halo stars will ultimately be powerful constraints upon the evolution
of supernova physics as a function of metallicity.

\subsection{Possible Modifications of the Basic Model}
\label{modifications}

As is clear from figures \ref{fig:WWsolopf} and \ref{fig:jnkpf}, the
simplest case of a non-magnetic non-rotating NDW from a neutron star
without additional energy deposition does not produce r-process nuclei in
significant abundances.  Are there extensions to this simple scenario
that {\sl could} make the wind a site of the r-process?

As was pointed out by \cite{Metzger07}, the combination of rotation
and magnetic fields can decrease the dynamical time scale by magnetic
``flinging''.  This is not particularly effective.  Adding a
non-thermal source of kinetic energy means that less thermal energy
must be put into the wind for it to escape the potential well.
Therefore, lower entropies are achieved.  It seems unlikely that this
mechanism, by itself, will salvage the NDW as a site for the full
r-process.  If there were a way to make the rotation rate of the PNS
high enough, it might be possible that there would be a centrifugally
driven outflow.  Then the electron fraction would be determined by kinetic
equilibrium much deeper in the PNS envelope, and the material in the
outflow would have an electron fraction much lower than that seen in
the wind.

To test this possibility, we ran calculations with a centrifugal force
term added and corotation with the PNS enforced out to $10^3$ km.
Unfortunately, for reasonable PNS spin rates (20 ms period), 
we found this had little effect on the nucleosynthesis.  These
calculations were in a regime were the electron fraction was still set
by neutrino interactions.

Many authors have discussed the possible effects of both
matter-enhanced\citep{Qian95,Sigl95} and collective neutrino
\citep{Pastor02,Duan06} oscillations on NDW nucleosynthesis.  If
electron antineutrinos could undergo a collective oscillation near
the launch radius while the electron neutrinos did not, this would
increase the average energy of the antineutrinos if the $\mu$
and $\tau$ neutrinos have a significantly higher temperature,
facilitating a reduction in the electron fraction.  For a normal mass hierarchy however,
matter enhanced neutrino oscillations would probably cause electron
neutrino flavor conversion, which would {\sl increase} the electron fraction
and decrease the probability of significant r-process nucleosynthesis
\citep{Qian95}.

Collective neutrino oscillations can cause antineutrino oscillations
in the region were the electron fraction is set, and thereby decrease
the electron fraction where pure MSW oscillations would have predicted
an increased electron fraction\citep{Duan06}.  Clearly, the main
effect of oscillations would be on the composition of the wind, not
the dynamics.  As can be seen in the neutrino two color plots,
oscillations would have to change the effective temperature of the
anti-neutrinos by a very large amount to move from a region where N=50
close shell nucleosynthesis occurs to a region where the second
r-process peak can be produced.

These effects are based upon the assumption that $\mu$- and
$\tau$-neutrinos are significantly more energetic than the electron 
neutrinos.  In the calculation
of \cite{Woosley94}, this is the case, as can be seen in figure
\ref{fig:woos94_neut}.  Interestingly, the $\mu$ and $\tau$
temperatures are almost the same as the electron anti-neutrino
temperature in the \cite{Huedepohl09} calculation, which can be seen
in figure \ref{fig:jnk_neut}.  It is not clear wether this difference
obtains because of the difference in the PNS masses or the
significantly different neutrino physics employed in the calculations.
A detailed study of neutrino transport in static backgrounds showed
that the inclusion of all relevant neutrino interactions brings the
average energies of the $\mu$- and $\tau$- neutrinos closer to the
temperature of the anti-electron neutrinos \citep{Keil03}.  Therefore
it is uncertain wether or not neutrino oscillations could effect
nucleosynthesis significantly.  Clearly, the uncertainties here are
not in the wind itself but in the formation of the spectra in the PNS
and the details of neutrino transport with neutrino oscillations.

Finally, it has been suggested \citep{Qian96,Nagataki05} that adding a
secondary source of volumetric energy deposition can significantly
increase the entropy of the wind, which results in a more alpha-rich
freeze out and conditions that would be more favorable for r-process
nucleosynthesis.  The addition of energy to the wind also decreases
the dynamical timescale.  Since the important quantity to consider for
the r-process is $s^3/\tau_d$ \citep{Hoffman97}, 
both effects increase the chance of having a significant neutron to
seed ratio after freeze out.  If the NDW model is to be salvaged, this
seems to us the minimal necessary extension.  Of course, the physical
process contributing this extra energy is very uncertain.  One
possibility is that oscillations of the PNS power sound waves
which produce shocks and deposit energy in the wind, similar to the supernova
mechanism of \cite{Burrows06}, but smaller in magnitude.  We will
explore this possibility in some detain in a subsequent paper.

From a chemical evolution standpoint, it is important to consider what
effect neutron star mergers will have on the evolution of the
r-process abundances.  It seems unavoidable that r-process nuclei will
be produced in the tidal tails ejected during these mergers
\citep{Freiburghaus99} and the amount of material ejected in these
events is approximately enough to account for the galactic inventory
of r-process elements given the expected merger rate
\citep{Lattimer76,Rosswog99}.

Neutron star mergers have been largely discounted because inferred
merger rates are small at low metallicity due to the long in spiral time
and therefore they cannot account for the r-process enrichment seen in
low metallicity halo stars \citep{Argast04}.  Of course, the inferred
merger rates are very uncertain as are models for the early evolution
of the Milky Way, so that both the conclusion that neutron star
mergers can account for the r-process inventory of the galaxy and that
they are not consistent with producing the r-process at low
metallicity are very uncertain.  Clearly, there is significant room
for more work in this area.

Therefore, it seems reasonable that the galactic r-process abundances
could be accounted for by a combination of mergers and winds with an
extra source of energy, either from an acoustically and/or magnetically
active PNS.  Since not every supernova will have the requisite
conditions for an r-process, there will be significant variation in
the yields from single supernovae.  This, along with the contribution
from neutron star mergers, will give significant variation in the
[r-process/$\alpha$-element] values found in single stars at low
metallicity but averaged over many stars these should track one
another, which is consistent with observations \citep{Sneden08}.

\section{Conclusions}

We have performed calculations of the dynamics and nucleosynthesis in
time dependent neutrino driven winds.  This was done for two sets of
neutrino spectra calculated in one-dimensional supernova models taken
from the literature.  The nucleosynthesis in these models was compared
with supernova yields to determine if these models were consistent
with observations.  Additionally, we compared the results of these
numerical models to analytic models of the neutrino driven wind and
found good agreement.

Similar to most of the work on the NDW after \cite{Woosley94}, we find
that it is unlikely that the r-process occurs in the neutrino driven
wind unless there is something that causes significant deviation from
a purely neutrino driven wind.  Additionally, in the simplest case,
there is little production of p-process elements at early times in the
wind.  In our calculation that used spectra from a more massive
neutron star, the wind only produces the N=50 closed shell elements
$^{87}$Rb, $^{88}$Sr, $^{89}$Y, and $^{90}$Zr.

This result is sensitive to small changes in the neutrino interaction rates 
(i.e. the inclusion of weak magnetism) and changes to the neutrino temperature 
of order 10\%.  Comparing our models with the over abundance of strontium 
seen in SN 1987A suggests that the difference between the electron and 
anti-electron neutrino temperatures in the model of \cite{Woosley94} 
may have been to large.  We also find that the effect of a wind termination
shock on the wind nucleosynthesis is small.

Using neutrino spectra from an $8.8 M_\odot$ supernova that drives a
wind which is proton rich throughout its duration \citep{Huedepohl09}, we
find that no significant $\nu$p-process occurs and the wind does not
contribute to the yields of the supernova. The neutrino spectra from
this model are probably more accurate than the spectra from the model
of \cite{Woosley94}.  We also investigated
the effect of an outer boundary pressure which resulted in a wind 
termination shock.  This had a negligible effect on the nucleosynthesis.

However, one also expects that the nucleosynthesis in the NDW will
vary considerably from event to event, especially with the mass and
possibly the rotation rate of the PNS. The winds from more massive PNS
have greater entropy and might, in general, be expected to produce
heavier elements and more of them. The neutrino spectral histories of
PNS as a function of mass have yet to be determined over a wide 
range of parameter space.  Currently, the neutrino luminosities and 
temperatures are the largest uncertainties in models of the NDW.

\begin{acknowledgements}

We would like to thank Alex Heger, David Lai, Enrico Ramirez-Ruiz,
Sanjay Reddy, and Yong-Zhong Qian for useful discussions about issues
relating to this work.  L. R. was supported by an NNSA/DOE Stewardship
Science Graduate Fellowship (DE-FC52-08NA28752) and the University of
California Office of the President (09-IR-07-117968-WOOS).  S. W.  was
supported by the US NSF (AST-0909129), the University of California
Office of the President (09-IR-07-117968-WOOS), and the DOE SciDAC
Program (DEFC-02-06ER41438).  R. H. was supported by the DOE SciDAC
Program (DEFC-02-06ER41438) and under the auspices of the Department of 
Energy at Lawrence Livermore National Laboratory under contract 
DE-AC52-07NA27344.
\end{acknowledgements}

\appendix

\section{Analytic Wind Dynamics}
To understand the wind dynamics, we follow arguments similar to those 
of \cite{Qian96} and \cite{Cardall97}.  Conservation of the stress-energy 
tensor and number flux in a Schwarzschild geometry in steady state leads 
to the wind equations in critical form 
\begin{equation}
\label{eq:nconserve}
\dot M = 4 \pi r^2 m_b n \gamma y v
\end{equation}
\begin{equation}
\label{eq:bernoulli}
\gamma y v \td{}{r} \ln(\gamma y h_r) = \frac{\dot q m_b}{c^2 h_r} 
\end{equation}
\begin{equation}
\label{eq:momentum}
\gamma^2 \left(v^2 -c_s^2\right) \td{v}{r} = \frac{v}{r} \left[ 2 c_s^2 
- \frac{G M}{r y^2} \left(1 - \frac{c_s^2}{c^2} \right)\right]
 - \frac{\dot q m_b}{3 \gamma y h_r}
\end{equation}
here $\dot M$ is the rest mass loss rate, $m_b$ is the baryon mass, $n$ 
is the baryon number density, $v$ is the velocity measured by an observer 
at rest in the Schwarzschild frame, $\gamma = \left(1-v^2/c^2 \right)^{-1/2}$, 
$\dot q$ is the neutrino heating rate per mass, $c_s$ is the sound speed, $M$ 
is the neutron star mass, $h_r = 1 + \epsilon/(m_b c^2) + P/(nm_bc^2)$, $P$ is the pressure, 
and $\epsilon$ is the energy per baryon not including the rest mass.  To fully describe 
the wind, these equations must be supplemented by a set of nuclear rate 
equations and an equation of state.  For our analytic calculations we 
used a radiation-dominated non-degenerate equation of state comprised 
of relativistic electrons, positrons, and photons.

First, we estimate where the critical radius (i.e. the radius where radiation 
pressure equals the nucleon gas pressure) sits 
in relation to the neutrino sphere.  Neglecting the temperature
gradient in the equation of hydrostatic balance gives the density
structure of the atmosphere,
\begin{equation}
\log(n(r)/n_c) \approx -\int_{r_l}^{r} dr \frac{GMm_b}{r^2 y^2 T(r)}  
\end{equation}
The optical depth of this atmosphere for neutrinos is
\begin{equation}
\tau  = \sigma_\nu(\epsilon_\nu) \int_r^\infty \frac{n}{y} 
\approx \sigma_\nu(\epsilon_\nu) n_c  \int_r^\infty \frac{dr}{y} 
\exp\left(-\int_{r_l}^{r} dr' \frac{GMm_b}{r'^2 y'^2 T(r')}  \right)  
\end{equation}
Taking the neutrinosphere to be at an optical depth of $2/3$, the 
gas pressure equal to the radiation pressure at the critical radius, and 
approximating gravity as constant throughout the envelope, we arrive at 
an equation for the critical radius
\begin{eqnarray}
\label{eq:rcrit}
r_c &\approx& R_\nu \left[ 1 + \frac{T_c}{g m_b R_\nu} 
\ln \left( \frac{2}{3 \tau_0} y_\nu \right)\right]
\end{eqnarray}
where $\tau_0 = \sigma_\nu(\epsilon_\nu) n_c T_c/g m_b$.  For
characteristic values of $L_\nu$, $\epsilon_\nu$, $R_\nu$, and $M$,
$r_c$ is only a few percent larger than $R_\nu$.  This implies that
the GR corrections to the neutrino interaction rates at $r_c$ will be
at most a few percent.  For characteristic values, the GR correction
to gravity will be $y(r_l)^{-2} \approx y_\nu^{-2} \approx 1.5 $.
This agrees with the observation made by previous authors that GR
corrections to gravity dominate over corrections to the neutrino
interaction rates \citep{Cardall97,Thompson01}.

Assuming that most neutrino heating occurs near $r_c$, the entropy 
can be considered constant once the temperature cools to the nucleon 
recombination temperature ($T \approx 0.5$ MeV).  Therefore, 
the final nuclear abundances in the wind depend mainly on the 
wind entropy, electron fraction, and the timescale for outflow 
\citep{Qian96,Hoffman97}.  To determine the contribution of the 
wind to the nucleosynthesis of the entire supernova, the mass loss 
rate must also be known.  We now find estimates for these quantities 
and for the transonic radius of the wind.

To determine the asymptotic entropy, the total energy deposition 
per baryon needs to be estimated.  Using equation \ref{eq:bernoulli} 
and assuming the asymptotic velocity is small, the total energy 
deposited per baryon is 
\begin{equation}
\ln(\gamma_f)- \ln(y_c h_c) \approx - \ln(y_c h_c) 
\approx \int^\infty_{r_\nu} dr \frac{\dot q m_b}{\gamma y v c^2 h_r} 
\approx \int^\infty_{r_c} dr \frac{\dot q_\nu m_b}{\gamma y v c^2 h_r} 
= Q/(m_b c^2)
\end{equation} 
Considering that most of the neutrino energy is deposited near the 
hydrostatic atmosphere, the final entropy per baryon is approximately
\begin{equation}
\label{eq:ent}
s_f = \int^\infty_{r_\nu} dr \frac{\dot q m_b}{\gamma y v T } 
\approx \int^\infty_{r_c} dr \frac{\dot q_\nu m_b}{\gamma y v T} 
+ s_c \approx - \frac{m_b c^2 \ln(h_c y_c) h_{r,c}}{T_{c}} + s_c
\end{equation}
Assuming that the neutron-proton rest mass difference is negligible, 
the entropy of the envelope is negligible, and taking the relativistic 
enthalpy outside the logarithm to be one results in the scaling relation
\begin{equation}
s_f \approx  464 \ln\left(y_c^{-2}\right) R_{\nu,6}^{1/3} 
L_{\nu,51}^{-1/6}\epsilon_{\nu,MeV}^{-1/3} 
\left(\frac{y_c}{y_\nu}\right)^{1/3}.
\end{equation}
Notice that all the general relativistic corrections, both due to gravity 
and to the neutrino interaction rates, increase the entropy from the 
non-relativistic case \citep{Cardall97,Otsuki00}.  As was discussed 
above, the dominant correction to the newtonian case is from the 
GR correction to gravity.

To fix the mass loss rate, we must estimate the velocity at the critical 
radius.  Taking the momentum equation in critical form and assuming 
approximate hydrostatic equilibrium and subsonic velocities gives  
$c_s^2 dv/dr \approx \frac{\dot q}{3 \gamma y}$. 
Assuming that acceleration has occurred over a scale height, we have
\begin{equation}
y v_c \approx \frac{h_{eff} \dot q (r_c)}{6 c_s^2}
\end{equation}
where the heating rate at the critical radius is divided by two to 
account for the fact that beneath the critical radius the net heating 
goes to zero over approximately a scale height so that a characteristic 
value of the heating rate is one half the heating rate at the critical 
radius.  The scale height is given by 
$h^{-1}_{eff} = (\rho + P) G M/Pr^2 y^2$.  Combining these 
with equation \ref{eq:nconserve} results in a mass loss rate of 
\begin{eqnarray}
\label{eq:mdot}
\dot M &\approx& 4 \pi r_c^4 \frac{P}{\rho+P}
y_c^2 \frac{m_b n_c \dot q_\nu(r_c)}{3 G M c_{s,c}^2}
\end{eqnarray}
Using our result for the entropy gives a scaling relation for the mass 
loss rate, 
\begin{equation}
\dot M \approx \sci{7.4}{-11} \, \textrm{M}_{\odot} \, \textrm{s}^{-1}
\, \frac{r_{l,6}^{4} 
L_{\nu,51}^{5/3} \epsilon_{\nu,MeV}^{10/3}}{R_{\nu,6}^{10/3}
\left(M/M_\odot\right) \ln\left(y_l^{-2}\right) } y_l^2 
\left(\frac{y_\nu}{y_l}\right)^{10/3}.
\end{equation}
 Notice that GR corrections reduce the mass loss rate significantly, 
 by about a factor of 2.  Only including GR effects on neutrino 
 propagation and energies decreases the mass loss rate by about $10\%$.  
 The reduced mass loss rate due to GR corrections does effect the 
 integrated nucleosynthetic yields, although the effect is not as great 
 as the effect of the GR corrections on the entropy. 

Given the entropy and the mass loss rate, we can solve for the 
evolution of temperature with radius outside of the heating 
region where the velocity is still small using the relation 
$\ln(h_r y) = \ln(y_c h_{r,c}) + Q \approx 0$, which gives
\begin{equation}
r \approx \frac{2 G M}{c^2} 
\left[1 - \left(1 + T s_f/m_b\right)^{-2} \right]^{-1}\\
\end{equation}
The dynamical timescale of the wind is defined by 
$\tau_d^{-1} = v \gamma y/n \left| dn/dr \right|$.  Steady 
state baryon number conservation yields
\begin{equation}
\frac{1}{n}\left| \td{n}{r} \right| =  \frac{2}{r} 
+ \frac{\gamma^2}{v} \td{v}{r}  + \frac{G M}{r^2 y^2}
\end{equation}
Neutrino energy deposition will no longer dominate 
the momentum equation when nucleons reform, but the 
velocity will be subsonic.  In this limit, the momentum 
equation yields
\begin{equation}
\frac{\gamma^2}{v}\td{v}{r} = \frac{GM}{r^2 y^2 c_s^2}
\left(1-c_s^2\right) - \frac{2}{ r}
\end{equation}
which results in an estimate for the dynamical timescale 
far inside the sonic point,
\begin{eqnarray}
\label{eq:tau}
\tau_d^{-1}&\approx&
\frac{G \dot M M}{4 \pi r^4 n m_b y^2 c_s^2}
\end{eqnarray}
For seed formation we are interested in the dynamical 
timescale around 2 MeV.  Combining with our result for 
the temperature structure of the atmosphere we have
\begin{eqnarray}
\tau_{d,2} &\approx& 
\frac{16 \pi (G M m_b)^3 m_b K }{9 s_f^4 \dot M} y^2 \\
&\approx& 8.2 \, \textrm{ms} \, 
\left(\frac{M}{1.4 M_\odot}\right)^3 \left(\frac{s_f}{100}\right)^{-4} 
\left(\frac{\dot M}{10^{-5} M_\odot \, \textrm{s}^{-1}}\right)^{-1} y^2 
\end{eqnarray}
To agree with the definition of the dynamical timescale given 
in \cite{Qian96}, this should be multiplied by three as our definition of
$\tau_d$ differs slightly from the one used in \cite{Qian96}.  We note 
that all of the scaling relations above are equivalent to those of 
\cite{Qian96} in the non-relativistic limit. 

From this discussion, it is unclear if a transsonic wind will obtain.  A 
reasonable criteria for transsonic solutions is that $\rho v^2 + P$ at the 
sonic point is greater than the pressure behind the supernova shock, which is 
approximately given by equation \ref{eq:Pbound}.  Equation 
\ref{eq:momentum} along with equation of state for the wind can be 
combined to give the temperature and sound speed at the sonic radius in 
terms of the sonic radius and known quantities
\begin{equation}
T_s = \frac{3 G M_{NS}  m_b}{2 r_s s_f y_s^2 
\left(1+\frac{G M_{NS}}{2 r_s y_s^2 c^2}\right)}  
\end{equation}
This results in an implicit equation for the sonic radius
\begin{equation}
\dot M \approx r_s^{-3/2} \frac{16 \pi m_b^{1/2} K}{3^{3/2}  
s_f^{1/2}} y_s \gamma_s \left(\frac{3 G M_{NS} m_b}
{2 s_f y_s^2 \left(1+\frac{G M_{NS}}{2 r_s y_s^2 c^2}\right)}\right)^{7/2} 
\end{equation}
where $K$ is the radiation constant for photons and leptons combined.
In the non-relativistic limit, the sonic radius reduces to
\begin{equation}
r_s \approx 860 \, \textrm{km} \, \left( \frac{s_f}{100}\right)^{-8/3} 
\left(\frac{M_{NS}}{1.4 M_\odot}\right)^{7/3} \left(\frac{\dot M}
{10^{-5} M_\odot \, \textrm{s}^{-1}}\right)^{-2/3}.
\end{equation}
The wind termination shock position is approximately given by \citep{Arcones07}
\begin{equation}
R_{rs} \approx \sci{1.3}{3} \, \textrm{km} \left(\frac{\dot M}
{10^{-5} M_\odot \, \textrm{s}^{-1}}\right)^{1/2} \left(\frac{R_s}
{10^9 \, \textrm{cm}}\right)^{3/2} \left(\frac{E_{sn}}
{10^{51} \, \textrm{erg}}\right)^{-1/2} \left(\frac{v_w}
{10^{9} \, \textrm{cm s}^{-1}}\right)^{1/2}  	 
\end{equation}
where $E_{sn}$ is the supernova explosion energy, $R_s$ is the radius 
of the supernova shock, and $v_w$ is the wind velocity just inside the 
wind termination shock.

\section{Analytic Wind Nucleosynthesis}

The electron fraction in the wind is given by kinetic equilibrium of 
neutrino interaction rates at the critical radius, as the temperature has 
decreased enough that lepton capture is unimportant \citep{Qian96}
\begin{equation}
\label{eq:ye}
Y_{e,f} \approx \frac{\lambda_{\nu_e}}{ \lambda_{\nu_e} + \lambda_{\bar \nu_e}}
\end{equation}
where $\lambda_{\nu_e}$ is the neutrino capture rate per baryon and 
$\lambda_{\bar \nu_e}$ is the anti-neutrino capture rate per baryon.  After 
weak interactions cease and the temperature has decreased about 0.5 MeV, 
alpha particles form in the wind.  The initial alpha number fraction is
$Y_{\alpha,i} \approx Y_e/2$ for neutron-rich conditions and 
$Y_{\alpha,i} \approx 1/2 -  Y_e/2$ for proton-rich conditions.

In both proton and neutron-rich winds, the nucleosynthesis will be
characterized by the neutron to seed ratio in the wind.  For winds
with $Y_e>\approx 0.5$, alpha particles recombine into $^{12}$C by the
standard triple alpha reaction and then experience alpha particle
captures up to approximately mass 56 \citep{Woosley92}.  The slowest
reaction in this sequence is $^4$He($2\alpha$,$\gamma$)$^{12}$C, so
the total number of seed nuclei produced is equal to the number of
$^{12}$C nuclei produced.  The rate of alpha destruction is given by
\begin{equation}
\td{Y_\alpha}{\tau} \approx - 14 \rho_0^2 Y_\alpha^3 \lambda_{3\alpha}
\end{equation}
where $\rho_0=m_b n$ is the rest mass density and $\lambda_{3\alpha}$ is the rate of triple alpha, which
includes double counting factors.  The factor of 14 comes from assuming alpha 
captures stop at $^{56}$Ni.  In general, we define an abundance by
$n_i = n Y_i$, where $n_i$ is the number density of species $i$ and $n$ 
is the baryon density.  
Using our definition of the dynamical timescale, we have 
$d\tau \approx -\frac{\tau_d}{3 T} dT$.  Transforming $Y_\alpha$ to 
a function of temperature makes the integral given above trivial to solve.  
Using a rate for triple-alpha from \cite{Caughlan88} and assuming the reaction 
flow stops at $^{56}$Ni results in a seed abundance at the end of the 
$\alpha$-process given by
\begin{equation}
Y_s \approx \frac{1-Y_e}{28} \left(1 - \left[1 + \sci{1.4}{5} 
\tau_d s_f^{-2} (1-Y_e)^2 \right]^{-1/2}    \right)
\end{equation}
where $s_f^{-2}$ enters because $\rho_0 \propto T^3/s_f$ for a radiation dominated 
equation of state and the density enters to the second power.
 
Under proton-rich conditions, the $\nu$p-process has the potential to occur.  
This process is similar to the rp-process, except that long lived beta-decays are 
bypassed by (n,p) reactions.  An estimate for the integrated number of free 
neutrons produced is $\tau_d \lambda_{\bar \nu_e}(T_9 \approx 2)$, so that 
the neutron to seed ratio is 
\begin{equation}
\label{eq:nsp}
\Delta_n \approx \tau_d \lambda_{\bar \nu_e}(T_9 \approx 2)  
\frac{Y_p}{ Y_s}. 
\end{equation}
Here, $\lambda_{\bar \nu_e}(T_9 \approx 2)$ is the neutrino capture rate 
at the seed formation radius.  A similar relation is found in \cite{Pruet06}. 
Although it is hard to estimate its effect, we note that the presence of a reverse 
shock can significantly affect the $\nu$p-process nucleosynthesis, as passage
through the reverse shock slows the outward flow and rarefaction of the wind.
Additionally, it increases the temperature to close to the post supernova 
shock temperature.  At early times for characteristic
explosion energies, the wind is shock heated to a temperature of a few GK.  
This all combines to give a longer period of time over which proton capture 
on heavy nuclei is efficient and allows the $\nu$p-process to continue to higher
mass than it would if no wind termination shock were present.

In the neutron-rich case, seed nuclei are produced by the slightly different 
reaction sequence $^4$He($\alpha$n,$\gamma$)$^9$Be($\alpha$,n)$^{12}$C 
\citep{Woosley92}.  For the conditions encountered in the wind, neutron 
catalyzed triple-alpha proceeds about ten times as quickly as 
$^4$He($2\alpha$,$\gamma$)$^{12}$C.  This implies that there will be a 
larger seed number than in proton-rich conditions.  The rate of helium 
destruction is given by the equations \citep{Hoffman97}
\begin{eqnarray}
\td{Y_\alpha}{\tau} \approx - \frac{\bar Z}{2} \rho_0 Y_\alpha 
Y_9 \lambda_{\alpha,n}(^9\textrm{Be})  \\
\td{Y_n}{Y_\alpha} \approx \frac{2(\bar A - 2 \bar Z)}{\bar Z} \\
Y_9 = \frac{27}{32}N_a^2 \left(\frac{2 \pi \hbar^2}{m_b}\right)^{3} 
\, Y_n Y_\alpha^2 \rho_0^2  T^{-3}  \exp((B_9-2 B_\alpha)/T) 
\end{eqnarray}
where $\bar Z$ is the average proton number of the seed nuclei, $\bar A$ 
is the average nucleon number of the seed nuclei, $N_a$ is Avogadro's 
number, $Y_\alpha$ is the alpha particle abundance, $Y_n$ is the neutron 
abundance, $Y_9$ is the $^{9}$Be abundance, $\lambda_{\alpha,n}(^9\textrm{Be})$
is the rate of $^{9}$Be destruction by $(\alpha,n)$, and $B_\alpha$ and $B_9$ 
are the binding energies of $^{4}$He and $^{9}$Be, respectively.  For 
$\lambda_{\alpha,n}(^9\textrm{Be})$, we employ the rate given in 
\cite{Wrean94}.  This set 
of equations can be solved analytically by once again transforming from 
proper time to temperature using the dynamical timescale.  The resulting 
implicit expression for the final alpha fraction is somewhat cumbersome, 
so we do not reproduce it here.  The seed abundance in terms of the initial 
and final alpha fraction is 
\begin{equation}
Y_s = 2 \frac{ Y_{\alpha,i}-Y_{\alpha,f}}{\bar Z}
\end{equation}
the final neutron fraction is given by 
\begin{equation}
Y_{n,f} = (1-2 Y_e) - 2 \frac{\bar A - 2 \bar Z}{\bar Z} 
\left( Y_{\alpha,i}-Y_{\alpha,f} \right)
\end{equation}
so that the neutron to seed ratio is 
\begin{equation}
\label{eq:nsn}
\Delta_n = \frac{\bar Z (1/2 - Y_e)}{Y_e/2-Y_{\alpha,f}} 
+ 2 \bar Z -\bar A.
\end{equation}
The seed abundance at the end of charged particle reactions can be 
estimated by \citep{Hoffman97}
\begin{equation}
Y_s \approx \frac{1-2Y_e}{10} \left( 1 - \exp\left[-\sci{8}{8}
 \tau_d s_f^{-3} Y_e^3\right] \right) 
\end{equation}
where it has been assumed that the neutron abundance is what 
limits the reaction.  Notice that the seed abundance in the proton-rich 
case depends on the entropy squared because it is mediated by an effective 
three body reaction, but in the neutron-rich case the entropy enters to 
the third power because of the effective four body interaction that 
mediates seed production.



\end{document}